# Will enterprise digital transformation affect diversification strategy?

*Ge-zhi Wu[1,2] | Da-ming You[1,2]*

[1] Business School, Central South University, Changsha, China

[2] Collaborative Innovation Center of Resource-conserving and Environmentally friendly Society and Ecological Civilization, Central South University, Changsha, China

**Correspondence**
Ge-zhi Wu, Business School, Central South University; Collaborative Innovation Center of Resource-conserving and Environmentally friendly Society and Ecological Civilization, Central South University, No. 932 Lushan South Road, Yuelu District, Changsha, Hunan 410083, China
　　Personal Email: 649792656@qq.com
　　Institutional Email：wgzz2218@csu.edu.cn
　　Telephone number: 86-18975833865

**Funding information**
　　National Natural Science Foundation of China, Grant/Award Numbers: 71974209;

**ACKNOWLEDGMENTS**
The authors acknowledge the anonymous reviewers and editors for helpful guidance on prior versions of the article.

## ABSTRACT

This paper empirically examines the impact of enterprise digital transformation on the level of enterprise diversification. It is found that the digital transformation of enterprises has significantly improved the level of enterprise diversification, and the conclusion has passed a series of robustness tests and endogenous tests. Through mechanism analysis, we find that the promotion effect of enterprise digital transformation on enterprise diversification is mainly realized through market power channel and firm risk channel, the pursuit of establishing market power, monopoly profits and challenge the monopolistic position of market occupiers based on digital transformation and the decentralization strategy to deal with the risks associated with digital transformation are important reasons for enterprises to adopt diversification strategy under the background of digital transformation. Although the organization costs channel, transaction costs channel, block holder control channel, industry type and information asymmetry channel have some influence on the main effect of this paper, they are not the main channel because they have not passed the inter group regression coefficient difference test statistically.

## 1| INTRODUCTION

The impact of information technology on the world has deepened day by day. A series of emerging information technologies and digital technologies, including big data and the Internet, have developed rapidly, which has had a far-reaching impact on the development of the global economy(Fitzgerald et al. 2014; Ross et al. 2016). Digital transformation has become an important engine of global economic growth(Agarwal et al. 2010; Majchrzak et al. 2016). Worldwide, enterprises in various countries gradually take digital technology as an important strategic measure of enterprise development, and upgrade and transform the organizational structure and business process through information technology(Matt et al. 2015).

In view of the tremendous impact of enterprise digital transformation, enterprises must design ways to maintain competitiveness, because digital technology "both game-changing opportunities and existential threats to companies." (Sebastian et al. 2020).With the rapid development of digitization, it is very important to understand the impact of enterprise digital transformation on enterprise strategic choice. It will not only help to further understand the enterprise strategy itself and future development trend, but also help to have an insight into the overall macro and global economic trend and development. It will be possible to provide strong policy suggestions for relevant managers. Therefore, it has important practical significance.

For many years, enterprise digitization has been the focus of research. Including enterprise strategy (bharadwaj et al. 2013; matt et al. 2015), organizational structure (Selander and jarvenpaa 2016), process (Carlo et al. 2012) and culture (Karimi and Walter 2015) and other aspects to study the impact of enterprise digital transformation on enterprise value contribution (svahn et al. 2017).

However, no research has empirically tested the impact of digital transformation on enterprise diversification. First, the improvement of digitization helps enterprises reduce organizational costs, further improve information processing efficiency, and finally promote the expansion of organizational scale (Malone et al., 1987), which will promote the diversified development of enterprises; Second, the technological advantages brought by the digital transformation of enterprises may help enterprises to adopt diversified strategies to obtain monopoly profits or as a weapon to challenge market rulers; Third, enterprise digital decision-making may help to improve the accuracy and breadth of decision-making of external major shareholders, and then help to improve the level of diversification. Fourth, due to the foresight and complexity of digital enterprises, enterprise digital transformation itself may aggravate the enterprise risk level, and digital technology itself will also enhance the ability of enterprises to use diversified strategies to resolve risks, which may promote enterprises to adopt diversification strategies to reduce enterprise risks.

On the other hand, the development of digitization also makes the information communication between enterprises and the outside world more border, reduces the transaction costs of enterprises(Malone et al., 1987), is conducive to promoting the specialized development of enterprises, and thus relatively reduces the diversification level of enterprises. In addition, the implementation of the enterprise digital transformation strategy strengthens internal control, which will help to prevent the generation of enterprise agency problems, which will curb the management's behavior of making profits for itself by adopting diversification strategy to build an empire. Overall, the impact of enterprise digital transformation on diversification needs further empirical test.

As for the reasons for selecting the main factors of the channel analysis path, we mainly consider that economies of scale, market forces and financial economics are considered as the main motivation for enterprises to adopt diversification strategies (Shi and hoskisson et al., 2021). Specifically, first, considering that economies of scale are one of the main motivation for enterprises to adopt diversification strategies (Rawley et al., 2010), we choose transaction costs and organizational costs. Second, considering that market power is the main factor affecting the diversification strategy of enterprises (Montgomery et al., 1985), we choose market power. Third, considering that financial economics is also an important factor affecting the diversification strategy of enterprises (Shi and hoskisson et al., 2021), we select the block holder shareholding and information asymmetry factors from the perspective of agency issues, and we select the risk factors closely related to financial risk management.

The marginal contribution of this paper lies in: At the theoretical level, this paper expands the theoretical boundary between diversification research field and digital research field. In particular, this paper extends the business ecosystem theory. The business ecosystem theory mainly focuses on organizational boundaries (Alexy et al., 2013; Meyer et al., 2005; Santos and Eisenhardt, 2005; Teece, 2007), characteristics (Zott and Amit, 2013), sub ecosystem (sub industry) ,system (Adner and Kapoor, 2010) and market participant behavior (viswanadham and samvedi, 2013), Among them, the digital ecosystem mainly focuses on the relationship between IT technology and enterprise business and market (Iyer et al., 2006; Selander et al., 2013). This paper further studies the changes of diversified strategic behaviors of enterprises as members and subjects of the business ecosystem in the digital transformation era, and it also helps to expand the understanding of the boundaries of enterprise diversification strategy, explain whether and how enterprise digital transformation affects diversification strategy, it provides theory and evidence for predicting the future changes of the whole business ecosystem. Firstly, Based on transaction cost theory, RBV, creative destruction, agency theory, DCV, this paper expands the research on the economic consequences of enterprise digitization, and empirically analyzes the impact of enterprise digitization transformation on enterprise diversification strategy **for the first time**, it found that the digital transformation of enterprises will increase the boundaries of enterprise diversification. Secondly, this paper studies the reasons why enterprises carry out diversification strategies in the digital context from multiple channels. Based on the transaction cost theory, we conduct research on transaction costs channel and organization costs channel; Based on RBV and creative destruction theory, we conduct market power channel research. Based on the agent theory, we conducted the research on block holder control channel and information asymmetry. Based on DCV theory, we study the firm risk channel. After empirical analysis, we finally explain the reasons why enterprises adopt diversified strategies in the context of digital transformation through market power channel and firm risk channel. Thirdly, this article helps to deepen the understanding of the essence of enterprise digital transformation, adds new literature to the influence of enterprise digital transformation, provides new driving factors and theoretical insights for the formulation of the diversification strategy, and expands the application of traditional RBV theory, creative destruction theory and DCV theory. At the practical level, examining the impact of enterprise digital transformation on enterprise diversification strategy from a micro perspective will help enterprise managers and relevant policy makers have an insight into the future development direction of enterprises and deeply understand the economic consequences of digital transformation.

## 2 | RELATED LITERATURE AND HYPOTHESIS DEVELOPMENT
### 2.1 TRANSACTION COSTS AND ORGANIZATION COSTS CHANNEL

Economies of scale and synergy are important reasons for enterprises to adopt diversification strategy (Penrose, 1959; Teece, 1980, 1981, 1982). However, due to the limitation of enterprise spatial distribution and geographical factors, it poses a challenge to enterprise management and increases the cost of enterprise management (Coase, 1937). Previous studies have shown that diversification strategy will lead to the dispersion of business interests (Tallman and Li, 1996). The dispersion of business interests further increases the requirements for information processing capacity, which will also aggravate the operating costs of enterprises (Morrison and Roth, 1992; Tallman and Li, 1996). transaction costs theory holds that the key to the boundary of enterprise diversification level is to balance the economic benefits brought by diversification to offset the resulting organization costs. (Jones and hill, 1988).

### 2.1.1 ORGANIZATION COSTS CHANNEL

Digital transformation can reduce the organization costs of enterprises: first, information technologies such as digital information management system can improve the organization's information processing ability and efficiency, facilitate the cooperation and communication between various divisions of the enterprise, realize low-cost recording and tracking of all matters in all links of supply, production and marketing, and optimize the coordination and linkage of all production links of the enterprise, Thus, the accuracy and precision of enterprise management decision-making are improved, It reduces the organization costs (Fernandez and Nieto, 2006). Second, the development of digital technology is conducive to the real-time and transparent recording of enterprise personnel management, production R & D, financial management and all important links, effectively reduces and compresses the possibility of agents engaging in default activities, thus reducing the agency cost of enterprises (Chen and Kamal, 2016). Therefore, the organization costs will decrease with the improvement of enterprise digitization, and the enterprise digital transformation will be conducive to the choice of enterprise diversification strategy.Based on this, hypothesis 1 is proposed:

H1：The digital transformation of enterprises will improve the diversification level of enterprises by reducing organizational costs.

### 2.1.2 TRANSACTION COSTS CHANNEL

Digital transformation also provides potential for enterprises to adopt professional strategies, which can reduce the transaction costs faced by enterprises: first, the development of digital technology can accelerate the collection, storage and analysis of information, help enterprises contact a wider range of business partners, and understand the potential capabilities, credit Transaction history and other information are helpful for enterprises to identify better counterparties, Reduces search costs for enterprises (Malone et al., 1987) second, digital technology deepens the communication between enterprises, makes the key subject information of transactions between enterprises more transparent, and reduces the negotiation cost in the process of signing enterprise contracts. Third, various technologies of digital mobile Internet and industrial Internet can ensure that the client can track the agent in time after signing the contract , This will lead to a significant reduction in the regulatory cost of enterprises under digitization (Clemons, 1993) in addition, the timely contact between enterprises and real-time tracking of materials brought by digital technology ensure that even if the contract is not listed, the transaction details can be flexibly adjusted between enterprises and counterparties according to immediate needs, so as to reduce the production related costs caused by incomplete contracts. Fourth, highly transparent transaction records under digital technology , the opponent will bear higher reputation loss in case of default, which further reduces the possibility of default. To sum up, the transaction costs will be reduced with the improvement of enterprise digitization, and the enterprise digital transformation will be conducive to the professional development of enterprises. Since the relationship between enterprise specialization and diversified development is often an opposite group, we believe that the development of enterprise specialization will weaken the possibility of enterprise diversified development.Based on this, hypothesis 2 is proposed:

H2：The digital transformation of enterprises will reduce the diversification level of enterprises by reducing transaction costs.

## 2.2 MARKET POWER CHANNEL

The traditional method of industrial economics holds that the operation of enterprises in multiple products, markets and businesses is in the pursuit of market monopoly power and associated benefits. Therefore, in the case of vertical integration, companies with a monopoly position at a certain stage of the value chain can use the market to foreclose and extend their monopoly to the adjacent stages of the value chain (chemow, 1998). The development of vertical integration forces competitors to take the way of vertical integration to check and balance, thus increasing entry barriers. Generally speaking, large diversified companies can exercise market power in three ways: 1 Predatory pricing. 2 Reciprocal buying. 3 Mutual forbearance.The digital transformation of an enterprise has a significant impact on the market in which the enterprise is located (mithas et al. 2013).

RBV believes that valuable, rare, incompletely imitatable and replaceable resources are an important source for enterprises to obtain sustainable competitive advantages (Barney, 1991; Coates and Mcdermott, 2002; Dyer and Singh, 1998). Specifically, if a resource can create new market opportunities or has the influence to eliminate the threat of competitors, it is valuable (Barney, 1991). The scarcity of resources indicates that resources among enterprises are heterogeneous, and competitors cannot obtain the same resources to create similar value (Coates and McDermott, 2002; Karim et al., 2022). If an enterprise has scarce resources, it can gain competitive advantages, but the duration depends on the impossibility and substitutability of resources. (Barney，1991）.Digital transformation technology obviously has the characteristics of the resources mentioned above.

On the one hand, according to RBV theory(Wernerfelt et al,1984), digital technology as a technical resource，the digital transformation of enterprises is likely to strengthen the technological monopoly power of enterprises and further consolidate the monopoly position of enterprises, which leads enterprises to adopt relevant diversification strategies to obtain monopoly profits and will establish new competitive and monopoly advantages.

Creative Destruction claims that innovation is both creation and destruction - the destruction of old methods and products ushers in the creation of new methods and products. The breakthrough technological innovation and the technology brought by it will cause great conflict to the industrial competition pattern, that is, the old industry will die out and the new industry will rise (Schumpeter, 1950).

On the other hand, From the perspective of the market as a whole，the digital transformation recombines existing products and services into new services (Yoo et al. 2010), supports services based on new products (Barrett et al. 2015), reducing barriers to entry (Woodard et al., 2013) and Hindering the sustainability of the competitive advantage of existing players (kahre et al2017). For example, platforms can redefine existing markets (Tiwana et al.2010), by promoting the exchange of digital goods and services. As the competition goes on, from the physical plane to the virtual plane with more free flow of information, previous form of entry barriers becomes less important, it consistent with the theory of creative destruction. We believe that the digital transformation of enterprises provides an opportunity for existing enterprises to challenge the market occupiers. Under the condition of high market concentration, enterprises will seek market breakthrough points through diversified strategies based on the digital transformation of enterprises, launch new products and services, break through the market with the opportunity of creative destruction, and finally enhance market power. Based on the above points of view, we believe that the pursuit of strengthening the monopoly power of the existing market and the pursuit of breakthroughs in the high monopoly market are the main reasons for enterprises to adopt diversified strategies in the context of digital transformation. Based on this, hypothesis 3 is proposed:

H3：The digital transformation of enterprises will enhance the diversification level of enterprises by enhancing their monopoly position or challenging the monopolistic position of market occupiers.

## 2.3 BLOCKHOLDER CONTROL CHANNEL

According to the agency theory, when the principal entrusts the agent with a task, because the interests of the two are inconsistent and there is no effective supervision mechanism to supervise the agent's behavior, the agent will do some opportunistic behaviors for his own interests, which may damage the interests of the principal(Jensen et al., 1976; Eisenhardt et al., 1989; Fama et al., 1983). In the case of asymmetric information, managers and large shareholders may use diversification strategies to build empire for their own interests, in order to consolidate the status of managers or erode the rights and interests of small shareholders.(Bosse et al.,2016)

Research on corporate diversification shows that block holders tend to promote diversification. According to hautz et al. (2013), ownership concentration is positively correlated with product diversification. Nguyen (2018) believes that the existence of block holders encourages the diversification of Vietnamese enterprise. Gu et al. (2018) believe that the non controlling shareholders of Chinese enterprises have strong supervision and incentive, which stimulates diversification. We believe that enterprise digital transformation has greatly improved the information and decision-making accuracy required for enterprise strategic decision-making, and may improve the accuracy of decision-making of major shareholders, according to the agency theory, this may encourage enterprises to use diversification for empire building, so as to improve the degree of enterprise diversification. Based on this, hypothesis 4 is proposed:

H4: The digital transformation of enterprises will improve the diversification level of enterprises by improving the decision-making level of major shareholders.

## 2.4 INFORMATION ASYMMETRY CHANNEL

External monitoring mechanisms play a key role in monitoring management performance and help alleviate diversification decisions that lead to the decline of enterprise value (Denis et al., 1997). Companies with higher levels of information disclosure generally have a lower level of diversification (Gu et al., 2018). According to agency theory, information symmetry tends to prevent diversified opportunistic management decisions and curb the emergence of agency problems(Eisenhardt et al.,1989). We believe that the digital transformation of enterprises involves the storage and packaging of information, which may greatly enrich the enterprise information environment, so as to strengthen the enterprise information disclosure, inhibit the generation of agency problems, and prevent managers from arbitrarily adopting diversification strategies for reasons such as Empire construction, so as to reduce the degree of enterprise diversification. Based on this, hypothesis 5 is proposed:

H5: The digital transformation of enterprises will reduce the level of enterprise diversification by reducing information asymmetry.

**2.5 FIRM RISK CHANNEL**

Diversification strategy is one of the important measures for enterprises to disperse risks(Gomez‑Mejia et al.,2010).Pandya and Rao (1998) found that diversified companies have lower risk and higher performance. Hitt et al. (1997) proved that diversified companies tend to perform better than non-diversified companies and have lower performance risk.The information advantage brought by digital transformation enables enterprises to integrate existing resources, find new business opportunities, and tend to choose unrelated diversification strategies to reduce risks(Woodard et al., 2013). Digitization is often closely related to the application of new technologies. One the one hand, the volatility of the company's performance and risk can be attributed to the application of new technologies Therefore, the digital transformation of enterprises may be transformed into the volatility of corporate performance and risk(Newell and Marabelli ,2015), so as to encourage the diversification of companies to reduce risks.

DCV explored the sources of wealth creation in the dynamic environment (Teece et al., 2016) and proposed that dynamic capabilities are the key environment for enterprises to gain competitive advantages in the changing environment (Helfat and Winter, 2011). The winners in the market are those companies that show timely response, rapid and flexible production innovation, and effective coordination of internal and external resources (Teece et al., 1997). Digital technology obviously has the above characteristics of improving the dynamic capability of enterprises.

One the other hand, according to DCV, Technological changes brought by digital transformation can increase the flexibility and emergency response capability of enterprises, the digital transformation of enterprises has also improved the risk-taking of enterprises(Tian et al.,2022), this increase in risk-taking ability is largely due to the more value growth channels brought by diversification strategy based on digital transformation.

Based on this, hypothesis 6 is proposed:

H6:The digital transformation of enterprises will enhance the diversification level of enterprises by increasing the risks of enterprises or improving enterprise risk-taking capacity.

## 2.6 ENTERPRISE DIGITAL TRANSFORMATION AND DIVERSIFICATION STRATEGY

Based on the above discussion, organization costs channel, market power channel, blockholder control channel and firm risk channel indicate that enterprise diversification transformation may improve the degree of enterprise diversification, while transaction costs channel and information asymmetry channel indicate that enterprise diversification transformation may reduce the degree of enterprise diversification, Then whether the enterprise diversification transformation can improve the enterprise diversification level depends on the size and direction of the comprehensive action of multiple forces.

Based on above discussion, the following assumptions H7 and H8 are put forward:

H7: When the net effect of enterprise digital transformation on enterprise diversification is positive, enterprise digital transformation promotes the diversified development of enterprises.

H8: When the net effect of enterprise digital transformation on enterprise diversification is negative, enterprise digital transformation inhibits the diversified development of enterprises.

## 3| SAMPLE DESCRIPTION AND VARIABLE MEASUREMENT

### 3.1 Data source and sample selection

This study collects relevant data from the databases of the China Stock Market & Accounting Research Database (CSMAR) and Chinese Research Data Services (CNRDS) and uses the main financial indicators and data listed on the Chinese A-share market from 2004 to 2020 as the initial sample. In the data collection process, we tried our best to ensure that the sample size was maximized. we eliminated the insufficient and missing data of related variables，. The final sample included 32907 company annual observations, representing 3709 unique companies.

Table 1 lists sample screening and sample distribution.

# Table 1 Sample screening and Sample distribution

**Panel A: Sample screening**

|  | observations |
|---|---|
| Initial sample | 46736 |
| Excluding: |  |
| Samples with missing and abnormal data | 13829 |
| Final sample | 32907 |

**Panel B: Sample distribution by year**

| Year | n | %_Total |
|---|---|---|
| 2004 | 431 | 1.31 |
| 2005 | 732 | 2.22 |
| 2006 | 888 | 2.70 |
| 2007 | 995 | 3.02 |
| 2008 | 1,126 | 3.42 |
| 2009 | 1,218 | 3.70 |
| 2010 | 1,409 | 4.28 |
| 2011 | 1,769 | 5.38 |
| 2012 | 2,178 | 6.62 |
| 2013 | 2,248 | 6.83 |
| 2014 | 2,247 | 6.83 |
| 2015 | 2,407 | 7.31 |
| 2016 | 2,643 | 8.03 |
| 2017 | 2,851 | 8.66 |
| 2018 | 3,190 | 9.69 |
| 2019 | 3,276 | 9.96 |
| 2020 | 3,299 | 10.03 |
| total | 32,907 |  |

**3.2| Variable description**

**3.2.1 Level of firm diversification**

We use two of the most commonly used measures of portfolio diversity to capture a firms level of diversification: Total entropy, Herfindahl index. Each measure is calculated using annual data from CSMAR. This study takes the classification standard of the industry classification guidelines of listed companies issued by China Securities Regulatory Commission as the main basis for the cross industry operation of listed companies

Total entropy captures the extent of diversity across a firm's activities (Jacquemin and Berry, 1979; Palepu, 1985). It is calculated as

$$Total\ entropy = \sum_{i=1}^{N} S_i \ln(1/S_i)$$

where Si is the share of a firm's total sales in industry i and N is the number of industries in which the firm operates. Total entropy equals zero for a single business firm and it rises with the extent of diversity.

The Herfindahl index of diversity is calculated as

$$\text{Herfindahl index} = \sum_{i=1}^{N} (S_i)^2$$

where Si is the share of a firm's total sales in industry i and N is the number of industries in which the firm operates. Since lower values of the Herfindahl index indicate higher levels of diversification We take a negative number for this value for comparative analysis. This measure equals -1 for a single business firm and it rises with the level of diversification.

**3.2.2 Digitization**

The quantitative measurement of enterprise digital transformation is a hot issue concerned by all parties. Enterprise digital transformation needs the help of cutting-edge digital technology and hardware system to promote the digitization of enterprise production process and means of production. First, the enterprise will focus on updating and upgrading the original technology and manufacturing system by relying on the "digital core technology drive". Among them, artificial intelligence, blockchain, cloud computing, big data and other technologies constitute the core underlying technology framework. The promotion of core and bottom technology focuses on the deep embedding of digital technology, mainly focusing on the digitization of core technologies and processes in various links such as internal management, production and operation of the enterprise; In addition, enterprise digital transformation is to form final output, product innovation and service innovation in the market. With the deepening of digital transformation, it will involve the core products and businesses of enterprises and form new performance growth points. At this level, technology pays more attention to the effective integration of digital and business environment. Based on this, in the structural framework of enterprise digital transformation, this research is divided into two levels: "bottom technology application" and "technology practical application": the "bottom technology application" includes four main technical components; In "technology practice application", focus on the application of specific digital services.

Follow the previous research methods(Jiang et al,2022;Tian et al,2022), we believe that as a major strategy for enterprise development in the period of enterprise digital transformation, it will be easier to be reflected in the enterprise annual report, which is summary, guiding and forward-looking. The use of words in the annual report can reflect the strategic choice characteristics and future vision of the enterprise, and to a large extent reflect the business philosophy and future development path of the enterprise. Therefore, it is scientific to measure the degree of digital transformation from the perspective of word frequency statistics involving "enterprise digital transformation" in the annual reports of listed enterprises.

This paper collects and arranges the annual reports of all A-share listed enterprises in Shanghai and Shenzhen through the python crawler function, and extracts all text contents through the Java pdfbox library as the data pool for subsequent feature word screening. Based on a series of classic literature(Westerman et al.,2011; Nwankpa and Roumani.,2016; Bekkhus., 2016; Haffke et al., 2016; Hartl and Hess.,2017;Paavola et al., 2017; Horlach et al., 2017; Legner et al., 2017; Liere-Netheler et al.,2018; Li et al., 2018), relevant important policy documents and research reports with the theme of digital transformation, this paper combs the key words of digital transformation, and finally confirms the characteristic words of digital transformation; The Related words are classified into "bottom technology application" and "technology practical application" to form a feature thesaurus. On this basis, the expressions of negative words such as "None" and "no" before deleting the keyword, and the "digital transformation" keywords unrelated to the company itself (including the introduction of company shareholders, customers, suppliers and company executives) have been deleted. Finally, based on the data pool formed by the annual report text extraction of Listed

Enterprises Based on python, search, match and count the word frequency according to the characteristic words, and then classify and collect the word frequency in the key technical direction to form the final aggregated word frequency, so as to build the index system of enterprise digital transformation. Because this kind of data has the typical characteristics of "skew distribution", this paper processes it logarithmically to obtain the overall index describing the digital transformation of enterprises. In the robustness test, this paper subdivides the caliber according to the composition difference and application status of the technology, and carries out the regression test again. See Appendix B for thesaurus.

**3.2.3 control variables.**

**Firm performance**

The company's performance is measured by the company's return on assets (ROA). ROA is a widely used performance measurement index. Relevant studies have shown that there is an important correlation between corporate performance and a number of other key indicators. (Keats and Hitt, 1988).For the same reason, we controlled the growth rate of the company's operating revenue, which is equal to (the company's operating revenue of the current year / the company's operating revenue of the previous year) - 1)

**Capital structure**

The research of O'Brien et al.(2014) based on agency theory shows that the company's capital structure will have an impact on diversification. Therefore, we use the proportion of the company's year-end liabilities to total assets to control this variable.

**Ownership structure**

Fox et al.(1994) research shows that the company's ownership structure will affect the diversification strategy. Therefore, we control the shareholding ratio of the management (the total shareholding of the management divided by the circulating share capital), the shareholding ratio of major shareholders (the number of shares held by the first major shareholder / the total number of shares), and the duality (if the chairman and general manager of the company are the same person, it is 1, otherwise it is 0). In addition, due to China's special ownership system, we also control whether the enterprise type is a state-owned enterprise or a private enterprise.

**Industry competitiveness**

Industrial competitiveness has been proved to be closely related to the degree of economies of scale and the degree of market power in the industry. Enterprises in highly competitive industries often show a lower level of diversification. (Christensen and Montgomery, 1981).This paper uses the huffindahl index, that is, the sum of the square of the proportion of the company's operating revenue in the operating revenue of all companies in the industry, to measure this variable.

**Industry capital intensity**

High industry capital intensity shows that production has a high degree of economies of scale, and high sunk costs form exit barriers, which affect the level of enterprise diversification.(Porter, 1980).We use the ratio of industry net fixed assets to industry employees to measure this variable.

**Firm size**

Firm size is a key index reflecting economies of scale and market power. Relevant studies show that there is a correlation between enterprise scale and diversification level (Grant, Jammine, and Thomas, 1988).We use the logarithm of the company's total assets to measure this variable.

**FirmAge**

The research of Xie et al(2014). Shows that corporate age will have an impact on diversification. Therefore, we control the logarithm of corporate age (current year - year of establishment + 1).

**3.3| The benchmark model**

To examine the relation between Digitization on diversification, we estimate the following equations :

$$Total\ entropy_{i,t}\left(Herfindahl\ index_{i,t}\right) = \beta_0 + \beta_1 Digitization_{i,t} + \sum \beta_j \left(Control\ variables\right)_{i,t} + \sum Firm_{i,t} + \sum Year + \varepsilon_{i,t}$$

$Total\ entropy_{i,t}$ and $Herfifindahl\ index_{i,t}$ refer to the diversification of company i in year t. Year and firm fixed effects are included to control for time- and firm-invariant factors.

The variables of interest, $Digitization_{i,t}$, measure the degree of digitization. A larger $\beta_1$ represents a greater impact of digitization on diversification. All variables are defined in Appendix A

# 4| EMPIRICAL RESULTS

## 4.1| Descriptive statistics and correlation analysis

Table 2-1 reports descriptive statistics of main variables for the 2004–2020 sample. Table 2-2 reports the Pearson correlation coefficients between variables. the correlation coefficients between independent variables and control variables were less than 0.5, So there is no evidence of severe multicollinearity among the variables.

**Table 2-1 Descriptive statistics results**

| VARIABLES | (1) N | (2) mean | (3) sd | (4) min | (5) max |
|---|---|---|---|---|---|
| Total entropy | 32907 | 0.407 | 0.473 | 0.000 | 2.620 |
| Herfindahl index | 32907 | -0.776 | 0.251 | -1.000 | -0.095 |
| $Digitization_{i,t}$ | 32907 | 0.885 | 1.218 | 0.000 | 6.043 |
| $Growth_{i,t}$ | 32907 | 5.303 | 747.094 | -1.000 | 134607 |
| $Roa_{i,t}$ | 32907 | 0.036 | 0.153 | -14.586 | 10.032 |
| $Lev_{i,t}$ | 32907 | 0.506 | 4.940 | 0.000 | 877.256 |
| $Mshare_{i,t}$ | 32907 | 0.115 | 0.255 | 0.000 | 22.567 |
| $FirmAge_{i,t}$ | 32907 | 2.801 | 0.381 | 0.693 | 4.143 |
| $Size_{i,t}$ | 32907 | 22.081 | 1.484 | 10.842 | 31.138 |
| $Tophold_{i,t}$ | 32907 | 0.344 | 0.152 | 0.003 | 0.900 |
| $Dual_{i,t}$ | 32907 | 0.249 | 0.432 | 0.000 | 1.000 |
| $Soe_{i,t}$ | 32907 | 0.393 | 0.488 | 0.000 | 1.000 |
| $HHI_{i,t}$ | 32907 | 0.104 | 0.114 | 0.015 | 1.000 |
| $CI_{j,t}$ | 32907 | 0.008 | 0.092 | 0.000 | 8.101 |

**Table 2-2 The correlation matrix of the explanatory variables**

| Variables | Total entropy | Herfindahl index | $Digitization_{i,t}$ | $Growth_{i,t}$ | $Roa_{i,t}$ | $Lev_{i,t}$ | $Mshare_{i,t}$ | $FirmAge_{i,t}$ |
|---|---|---|---|---|---|---|---|---|
| Total entropy | 1.000 | | | | | | | |
| Herfindahl index | 0.979a | 1.000 | | | | | | |
| $Digitization_{i,t}$ | 0.109a | 0.095a | 1.000 | | | | | |
| $Growth_{i,t}$ | 0.008 | 0.008 | -0.005 | 1.000 | | | | |

| Variables | Total entropy | Herfindahl index | Digitization$_{i,t}$ | Growth$_{i,t}$ | Roa$_{i,t}$ | Lev$_{i,t}$ | Mshare$_{i,t}$ | FirmAge$_{i,t}$ |
|---|---|---|---|---|---|---|---|---|
| Roa$_{i,t}$ | -0.037a | -0.039a | -0.001 | 0.001 | 1.000 | | | |
| Lev$_{i,t}$ | -0.000 | -0.001 | -0.011c | 0.000 | -0.070a | 1.000 | | |
| Mshare$_{i,t}$ | -0.118a | -0.115a | 0.131a | -0.003 | 0.063a | -0.015a | 1.000 | |
| FirmAge$_{i,t}$ | 0.106a | 0.101a | 0.145a | 0.004 | -0.041a | 0.008 | -0.113a | 1.000 |
| Size$_{i,t}$ | 0.250a | 0.194a | 0.137a | -0.002 | 0.019a | -0.036a | -0.165a | 0.197a |
| Tophold$_{i,t}$ | -0.053a | -0.052a | -0.122a | 0.006 | 0.062a | -0.010c | -0.066a | -0.156a |
| Dual$_{i,t}$ | -0.061a | -0.055a | 0.117a | -0.002 | 0.013b | 0.002 | 0.188a | -0.039a |
| Soe$_{i,t}$ | 0.130a | 0.117a | -0.184a | 0.006 | -0.027a | 0.004 | -0.343a | 0.055a |
| HHI$_{i,t}$ | 0.042a | 0.043a | -0.110a | -0.000 | -0.027a | 0.013b | -0.073a | -0.049a |
| CI$_{j,t}$ | 0.004 | 0.003 | -0.024a | -0.000 | -0.001 | 0.000 | -0.020a | 0.010c |

| Variables | Size$_{i,t}$ | Tophold$_{i,t}$ | Dual$_{i,t}$ | Soe$_{i,t}$ | HHI$_{i,t}$ | CI$_{j,t}$ |
|---|---|---|---|---|---|---|
| Total entropy | | | | | | |
| Herfindahl index | | | | | | |
| Digitization$_{i,t}$ | | | | | | |
| Growth$_{i,t}$ | | | | | | |
| Roa$_{i,t}$ | | | | | | |
| Lev$_{i,t}$ | | | | | | |
| Mshare$_{i,t}$ | | | | | | |
| FirmAge$_{i,t}$ | | | | | | |
| Size$_{i,t}$ | 1.000 | | | | | |
| Tophold$_{i,t}$ | 0.168a | 1.000 | | | | |
| Dual$_{i,t}$ | -0.136a | -0.057a | 1.000 | | | |
| Soe$_{i,t}$ | 0.271a | 0.239a | -0.279a | 1.000 | | |
| HHI$_{i,t}$ | 0.024a | 0.064a | -0.075a | 0.111a | 1.000 | |
| CI$_{j,t}$ | 0.036a | 0.009c | -0.018a | 0.041a | 0.063a | 1.000 |

This table reports pairwise Pearson correlation coefficients between the variables. a, b, and c indicate significance at the 1%, 5%, and 10% levels, respectively.

## 4.2| Multivariate analysis

### 4.2.1 Test of Hypothesis 7 and 8

According to the regression results in Table 3(1)-(4),enterprise digital transformation is positively correlated with the choice of diversification strategy, and significantly exists at the 1% confidence level, indicating that the net effect of enterprise digital transformation on enterprise diversification level is positive, and finally promotes the choice of enterprise diversification strategy. Hypothesis 7 is true, but hypothesis 8 is not true.

**Table 3 Digitization on diversification**

| Model | FE | | | |
|---|---|---|---|---|
| | (1) | (2) | (3) | (4) |
| Dependent variable | Total entropy | Total entropy | Herfindahl index | Herfindahl index |
| $Digitization_{i,t}$ | 0.0262*** | 0.0192*** | 0.0133*** | 0.0099*** |
| | (5.7459) | (4.1779) | (5.2502) | (3.8461) |
| $Growth_{i,t}$ | | 0.0000*** | | 0.0000*** |
| | | (17.8535) | | (20.7382) |
| $Roa_{i,t}$ | | -0.0170 | | -0.0111* |
| | | (-1.5612) | | (-1.7079) |
| $Lev_{i,t}$ | | 0.0002 | | -0.0000 |
| | | (1.4556) | | (-0.0618) |
| $Maghold_{i,t}$ | | -0.0207 | | -0.0118 |
| | | (-1.6156) | | (-1.4970) |
| $FirmAge_{i,t}$ | | 0.1463*** | | 0.0969*** |
| | | (3.4809) | | (4.1371) |
| $Size_{i,t}$ | | 0.0496*** | | 0.0236*** |
| | | (6.6697) | | (5.6840) |
| $Tophold_{i,t}$ | | -0.1326*** | | -0.0770*** |
| | | (-2.5823) | | (-2.6676) |
| $Dual_{i,t}$ | | 0.0018 | | 0.0009 |
| | | (0.2116) | | (0.1905) |
| $Soe_{i,t}$ | | 0.0283 | | 0.0135 |
| | | (1.3520) | | (1.1969) |
| $HHI_{i,t}$ | | -0.0236 | | -0.0011 |
| | | (-0.2991) | | (-0.0262) |
| $CI_{j,t}$ | | -0.0230** | | -0.0126** |
| | | (-2.2117) | | (-2.4296) |
| _cons | 0.4343*** | -0.8540*** | -0.7593*** | -1.4254*** |
| | (21.9201) | (-5.0108) | (-69.7993) | (-14.8964) |
| Firm FE | Yes | Yes | Yes | Yes |
| Year FE | Yes | Yes | Yes | Yes |
| Observations | 32907 | 32907 | 32907 | 32907 |
| adj. R2 | 0.035 | 0.050 | 0.033 | 0.046 |

This table reports the estimated results from the regressions of Digitization on diversification , t statistic based on the robust standard error is in parentheses.***, **, and * indicate significance at the 1%, 5%, and 10% levels, respectively.

**4.3| Robustness tests**

(1) In order to control the deviation of missing variables, we further added two control variables, industry R & D density $RD_{i,t}$ (proportion of industry R & D expenditure in total industry revenue) and industry export density $ED_{i,t}$ (proportion of industry export revenue in total industry revenue). Due to a large number of missing data in these two variables (sample size reduced from 32907 to 15799) and these variables do not have a significant impact on the model. In order to prevent sample selection bias, the relevant regression results are placed in the robustness test part, as shown in columns (1) - (2) of table 4-1. The main regression results are consistent with the benchmark regression.

(2) In order to increase the reliability of the research, this paper uses Tobit model to regress the main variables in this paper. It is worth mentioning that Tobit model can not control the firm fixed effect, so we only control the industry fixed effect. According to the results in table 4-1 (3) - (4) and Table 4-2 (1) - (2), the main regression results are consistent with the benchmark regression.

(3) We classify and regress several main components of digital technology keywords. According to the regression results in table 4-3 (1) - (5) and table 4-4 (1) - (5), we find that all technology categories are positively correlated with enterprise diversification strategy, and cloud computing, digital technology application and big data are significantly positively correlated with enterprise diversification strategy choice, These three types of digital technologies play a leading role in the choice of enterprise diversification strategy. We believe that the reason is that these three types of technologies are more mature and more available than artificial intelligence and blockchain technology, thus playing a key role in promoting the choice of enterprise diversification strategy.

**Table 4-1 Robustness tests of Digitization on diversification**

| Model | FE | | Tobit | |
|---|---|---|---|---|
| | (1) | (2) | (3) | (4) |
| Dependent variable | Total entropy | Herfindahl index | Total entropy | Herfindahl index |
| $Digitization_{i,t}$ | 0.0223*** | 0.0134*** | 0.0369*** | 0.0202*** |
| | (4.0267) | (4.1337) | (8.4606) | (8.3238) |
| $Growth_{i,t}$ | -0.0004*** | -0.0001 | -0.0001 | -0.0000 |
| | (-2.6623) | (-1.2330) | (-0.2109) | (-0.1786) |
| $Roa_{i,t}$ | -0.0748 | -0.0352 | -0.1333*** | -0.0753*** |
| | (-1.5968) | (-1.2181) | (-3.4960) | (-3.4829) |
| $Lev_{i,t}$ | -0.0111 | -0.0044 | -0.0227** | -0.0128** |
| | (-0.9562) | (-0.6212) | (-2.1037) | (-2.0835) |
| $Mshare_{i,t}$ | -0.0579 | -0.0398 | -0.1217*** | -0.0764*** |
| | (-1.3241) | (-1.4741) | (-4.2595) | (-4.7382) |
| $FirmAge_{i,t}$ | 0.0893 | 0.0612* | 0.2257*** | 0.1225*** |
| | (1.6196) | (1.9179) | (7.9568) | (7.8526) |
| $Size_{i,t}$ | 0.0656*** | 0.0369*** | 0.0938*** | 0.0496*** |
| | (4.8514) | (4.6929) | (15.1502) | (14.4115) |
| $Tophold_{i,t}$ | -0.1920** | -0.1229*** | -0.3885*** | -0.2128*** |
| | (-2.3777) | (-2.6287) | (-8.7346) | (-8.6187) |
| $Dual_{i,t}$ | -0.0060 | -0.0033 | -0.0153 | -0.0081 |
| | (-0.5757) | (-0.5447) | (-1.6067) | (-1.5329) |
| $Soe_{i,t}$ | -0.0033 | 0.0036 | 0.0277 | 0.0070 |
| | (-0.1210) | (0.2395) | (1.4432) | (0.6700) |
| $HHI_{i,t}$ | -0.0710 | -0.0115 | -0.1611 | -0.0868 |
| | (-0.5818) | (-0.1621) | (-1.5075) | (-1.4558) |
| $ED_{i,t}$ | -0.0689 | -0.0260 | -0.0720* | -0.0239 |
| | (-1.5290) | (-0.8166) | (-1.7675) | (-1.0280) |
| $CI_{j,t}$ | -1.2141 | -0.7694 | -0.7427 | -0.8505 |
| | (-0.8103) | (-0.9484) | (-0.5251) | (-1.0677) |
| $RD_{i,t}$ | -0.0043 | -0.0020 | -0.0052 | -0.0031 |
| | (-1.2017) | (-1.0503) | (-1.3095) | (-1.3861) |
| _cons | -1.1757*** | -1.6801*** | -2.4358*** | -2.2848*** |
| | (-3.8044) | (-9.3665) | (-9.4489) | (-16.2085) |
| Firm FE | Yes | Yes | | |
| INDUSTRYFE | | | Yes | Yes |
| Year FE | Yes | Yes | Yes | Yes |
| Observations | 15799 | 15799 | 15799 | 15799 |
| $R^2$ | 0.075 | 0.076 | | |

This table reports the estimated results from the regressions of robust test of Digitization on diversification. For the fixed effect model, t statistic based on the robust standard error is in parentheses, and for the Tobit model, z statistic based on the robust standard error is in parentheses. ***, **, and * indicate significance at the 1%, 5%, and 10% levels, respectively.

**Table 4-2 Robustness tests of Digitization on diversification**

| Model | Tobit | |
|---|---|---|
| | (1) | (2) |
| Dependent variable | Total entropy | Herfindahl index |
| $Digitization_{i,t}$ | 0.0250*** | 0.0134*** |
| | (7.9482) | (7.8781) |
| $Growth_{i,t}$ | 0.0000 | 0.0000 |
| | (0.7843) | (1.0721) |
| $Roa_{i,t}$ | -0.0304** | -0.0195** |
| | (-2.1847) | (-2.4944) |
| $Lev_{i,t}$ | 0.0000 | -0.0001 |
| | (0.0246) | (-0.3300) |
| $Maghold_{i,t}$ | -0.1867*** | -0.1049*** |
| | (-8.1719) | (-8.3631) |
| $FirmAge_{i,t}$ | 0.2263*** | 0.1255*** |
| | (11.3728) | (11.7760) |
| $Size_{i,t}$ | 0.0696*** | 0.0351*** |
| | (19.7119) | (18.2885) |
| $Tophold_{i,t}$ | -0.2482*** | -0.1328*** |
| | (-8.9707) | (-8.8070) |
| $Dual_{i,t}$ | -0.0052 | -0.0032 |
| | (-0.7486) | (-0.8376) |
| $Soe_{i,t}$ | 0.0393*** | 0.0167*** |
| | (3.5966) | (2.8312) |
| $HHI_{i,t}$ | -0.0052 | -0.0093 |
| | (-0.1015) | (-0.3364) |
| $CI_{j,t}$ | -0.0346 | -0.0207 |
| | (-1.3111) | (-1.3994) |
| _cons | -1.5443*** | -1.7734*** |
| | (-9.8034) | (-21.0577) |
| Firm FE | Yes | Yes |
| Year FE | Yes | Yes |
| Observations | 32907 | 32907 |

This table reports the estimated results from the regressions of robust test of Digitization on diversification. For the Tobit model, z statistic based on the robust standard error is in parentheses. ***, **, and * indicate significance at the 1%, 5%, and 10% levels, respectively.

**Table 4-3 Robustness tests of considering digital technology differences**

| Model | FE | | | | |
|---|---|---|---|---|---|
| | (1) | (2) | (3) | (4) | (5) |
| Dependent variable | Total entropy | Total entropy | Total entropy | Total entropy | Total entropy |
| $CC_{i,t}$ | 0.0133** | | | | |
| | (1.9799) | | | | |
| $DTA_{i,t}$ | | 0.0234*** | | | |
| | | (4.2210) | | | |
| $AI_{i,t}$ | | | 0.0017 | | |
| | | | (0.2139) | | |
| $BT_{i,t}$ | | | | 0.0138 | |
| | | | | (0.9351) | |

| | (1) | (2) | (3) | (4) | (5) |
|---|---|---|---|---|---|
| $DT_{i,t}$ | | | | | 0.0171*** |
| | | | | | (2.6032) |
| $Growth_{i,t}$ | 0.0000*** | 0.0000*** | 0.0000*** | 0.0000*** | 0.0000*** |
| | (18.2321) | (18.5449) | (18.3843) | (18.3618) | (17.9109) |
| $Roa_{i,t}$ | -0.0168 | -0.0169 | -0.0169 | -0.0167 | -0.0166 |
| | (-1.5422) | (-1.5470) | (-1.5345) | (-1.5266) | (-1.5160) |
| $Lev_{i,t}$ | 0.0002* | 0.0002 | 0.0002* | 0.0002* | 0.0002 |
| | (1.6792) | (1.5614) | (1.7657) | (1.7594) | (1.5984) |
| $Maghold_{i,t}$ | -0.0204 | -0.0212 | -0.0207 | -0.0205 | -0.0203 |
| | (-1.5875) | (-1.6413) | (-1.6069) | (-1.6013) | (-1.6118) |
| $FirmAge_{i,t}$ | 0.1453*** | 0.1472*** | 0.1466*** | 0.1466*** | 0.1459*** |
| | (3.4475) | (3.5045) | (3.4748) | (3.4809) | (3.4693) |
| $Size_{i,t}$ | 0.0520*** | 0.0509*** | 0.0531*** | 0.0529*** | 0.0509*** |
| | (7.0014) | (6.8737) | (7.1577) | (7.1419) | (6.8451) |
| $Tophold_{i,t}$ | -0.1386*** | -0.1375*** | -0.1414*** | -0.1410*** | -0.1357*** |
| | (-2.6891) | (-2.6698) | (-2.7395) | (-2.7312) | (-2.6391) |
| $Dual_{i,t}$ | 0.0019 | 0.0021 | 0.0020 | 0.0019 | 0.0016 |
| | (0.2286) | (0.2530) | (0.2424) | (0.2265) | (0.1882) |
| $Soe_{i,t}$ | 0.0276 | 0.0290 | 0.0282 | 0.0280 | 0.0281 |
| | (1.3159) | (1.3849) | (1.3415) | (1.3298) | (1.3383) |
| $HHI_{i,t}$ | -0.0289 | -0.0229 | -0.0313 | -0.0325 | -0.0311 |
| | (-0.3668) | (-0.2911) | (-0.3971) | (-0.4121) | (-0.3948) |
| $CI_{j,t}$ | -0.0241** | -0.0239** | -0.0246** | -0.0244** | -0.0239** |
| | (-2.3164) | (-2.3132) | (-2.3535) | (-2.3396) | (-2.2748) |
| _cons | -0.8960*** | -0.8805*** | -0.9179*** | -0.9147*** | -0.8750*** |
| | (-5.2539) | (-5.1843) | (-5.3833) | (-5.3806) | (-5.1225) |
| Firm FE | Yes | Yes | Yes | Yes | Yes |
| Year FE | Yes | Yes | Yes | Yes | Yes |
| Observations | 32907 | 32907 | 32907 | 32907 | 32907 |
| adj. $R^2$ | 0.049 | 0.050 | 0.048 | 0.048 | 0.049 |

This table reports the estimated results from the regressions of Digitization on diversification, t statistic based on the robust standard error is in parentheses.***, **, and * indicate significance at the 1%, 5%, and 10% levels, respectively.

**Table 4-4 Robustness tests of considering digital technology differences**

| Model | FE | | | | |
|---|---|---|---|---|---|
| | (1) | (2) | (3) | (4) | (5) |

| Dependent variable | Herfindahl index | Herfindahl index | Herfindahl index | Herfindahl index | Herfindahl index |
|---|---|---|---|---|---|
| $CC_{i,t}$ | 0.0078** | | | | |
| | (2.0422) | | | | |
| $DTA_{i,t}$ | | 0.0116*** | | | |
| | | (3.8487) | | | |
| $AI_{i,t}$ | | | 0.0020 | | |
| | | | (0.4459) | | |
| $BT_{i,t}$ | | | | 0.0093 | |
| | | | | (1.1208) | |
| $DT_{i,t}$ | | | | | 0.0086** |
| | | | | | (2.3854) |
| $Growth_{i,t}$ | 0.0000*** | 0.0000*** | 0.0000*** | 0.0000*** | 0.0000*** |
| | (20.6039) | (20.7424) | (20.5433) | (20.5173) | (20.6134) |
| $Roa_{i,t}$ | -0.0110* | -0.0111* | -0.0110* | -0.0110* | -0.0109* |
| | (-1.6945) | (-1.6952) | (-1.6824) | (-1.6766) | (-1.6683) |
| $Lev_{i,t}$ | 0.0000 | 0.0000 | 0.0000 | 0.0000 | 0.0000 |
| | (0.1860) | (0.0638) | (0.2861) | (0.2918) | (0.1018) |
| $Maghold_{i,t}$ | -0.0116 | -0.0120 | -0.0118 | -0.0116 | -0.0116 |
| | (-1.4705) | (-1.5183) | (-1.4895) | (-1.4835) | (-1.4928) |
| $FirmAge_{i,t}$ | 0.0963*** | 0.0974*** | 0.0969*** | 0.0970*** | 0.0967*** |
| | (4.1026) | (4.1595) | (4.1288) | (4.1395) | (4.1263) |
| $Size_{i,t}$ | 0.0248*** | 0.0243*** | 0.0253*** | 0.0253*** | 0.0243*** |
| | (5.9995) | (5.8734) | (6.1358) | (6.1301) | (5.8710) |
| $Tophold_{i,t}$ | -0.0799*** | -0.0796*** | -0.0813*** | -0.0812*** | -0.0787*** |
| | (-2.7605) | (-2.7489) | (-2.8053) | (-2.8018) | (-2.7233) |
| $Dual_{i,t}$ | 0.0010 | 0.0011 | 0.0010 | 0.0010 | 0.0008 |
| | (0.2037) | (0.2273) | (0.2154) | (0.1990) | (0.1706) |
| $Soe_{i,t}$ | 0.0131 | 0.0138 | 0.0134 | 0.0133 | 0.0134 |
| | (1.1598) | (1.2284) | (1.1897) | (1.1753) | (1.1849) |
| $HHI_{i,t}$ | -0.0037 | -0.0009 | -0.0051 | -0.0058 | -0.0050 |
| | (-0.0862) | (-0.0214) | (-0.1189) | (-0.1378) | (-0.1176) |
| $CI_{j,t}$ | -0.0131** | -0.0131** | -0.0134** | -0.0133** | -0.0131** |
| | (-2.5388) | (-2.5470) | (-2.5669) | (-2.5576) | (-2.4930) |
| _cons | -1.4452*** | -1.4397*** | -1.4563*** | -1.4557*** | -1.4366*** |

|  | (-15.1408) | (-15.0956) | (-15.2605) | (-15.2953) | (-15.0118) |
|---|---|---|---|---|---|
| Firm FE | Yes | Yes | Yes | Yes | Yes |
| Year FE | Yes | Yes | Yes | Yes | Yes |
| Observations | 32907 | 32907 | 32907 | 32907 | 32907 |
| adj. $R^2$ | 0.045 | 0.046 | 0.044 | 0.044 | 0.045 |

This table reports the estimated results from the regressions of Digitization on diversification , t statistic based on the robust standard error is in parentheses.***, **, and * indicate significance at the 1%, 5%, and 10% levels, respectively.

## 5| ADDRESSING ENDOGENEITY CONCERNS

There may also be endogenous problems in this paper. Based on this, considering that enterprise digital transformation is a positive response to the continuous maturity of "ABCD" technology, it is an excellent quasi natural experiment for enterprises to gradually promote their own digital transformation in batches. This paper selects the multi-stage dual difference model (did) to further overcome the endogenous problem: by making two differences between the treatment group and the control group before and after the implementation of the digital transformation strategy, it can effectively eliminate the internal differences between individuals and the errors caused by the time trend unrelated to the experimental group, We can get the "net effect" of enterprise digital transformation on enterprise diversification. Accordingly, this paper constructs the following double difference model:The model of the difference-in-differences regression is specified as follows:

$$Total\ entropy_{i,t}\left(Herfindahl\ index_{i,t}\right) = \beta_0 + \beta_1(du_{i,t} \times dt_{i,t}) + \sum \beta_j \left(Control\ variables\right)_{i,t} + \sum Firm_{i,t} + \sum Year + \varepsilon_{i,t}$$

Where, $du$ is an individual dummy variable, $du = 1$ represents the group of enterprises undergoing digital transformation during the sample period, $du = 0$ represents the group of enterprises that have not undergone digital transformation. Further set the period dummy variable $dt$. If the company carries out digital transformation in the current year and subsequent years, it will be assigned as 1, otherwise it will be 0. $\beta_1$ reflects the change of enterprise diversification before and after the enterprise promotes the digital transformation, and is the parameter to be evaluated of the key variable in this paper. It should be pointed out that the double difference samples need to have sufficient observation values in several years before and after the policy change. Therefore, the samples selected in this paper are samples with a period of at least five consecutive years, so as to ensure that there is sufficient observation period after the difference as much as possible. At the same time, this paper will eliminate those samples that have been showing digital transformation keywords during the whole sample period. In addition, the model controls the fixed effect of firms.

According to the regression results in Table 5 (1) - (2), we find that the variables ($du_{i,t} \times dt_{i,t}$) are still positively correlated with the enterprise diversification strategy, and significantly exist at the 5% confidence level, indicating that this study does not need to worry about the impact of endogenous problems on the conclusion of this paper. In addition, according to the regression results listed in Table 5 (3) - (4), we can see that the regression results of relevant variables are not significant three years ($du_{i,t} \times Before3_{i,t}$), two years ($du_{i,t} \times Before3_{i,t}$), one year ($du_{i,t} \times Before1_{i,t}$) before the implementation of the policy. One year ($du_{i,t} \times After1_{i,t}$), two years ($du_{i,t} \times After2_{i,t}$) and three years ($du_{i,t} \times After3_{i,t}$) after the implementation of the policy, the relevant results are positive and basically significant, which further illustrates the reliability of the benchmark regression results.

**Table 5 Difference-in-differences (DID) regressions.**

| Model | DID | | | |
|---|---|---|---|---|
| | (1) | (2) | (3) | (4) |
| Dependent variable | Total entropy | Herfindahl index | Total entropy | Herfindahl index |
| $du_{i,t} \times dt_{i,t}$ | 0.0245** | 0.0116** | | |
| | (2.3478) | (2.0093) | | |
| $du_{i,t} \times Before3_{i,t}$ | | | 0.0025 | 0.0011 |
| | | | (0.2780) | (0.2199) |
| $du_{i,t} \times Before2_{i,t}$ | | | -0.0002 | -0.0003 |
| | | | (-0.0223) | (-0.0648) |
| $du_{i,t} \times Before1_{i,t}$ | | | 0.0050 | 0.0029 |
| | | | (0.5350) | (0.5451) |
| $du_{i,t} \times current_{i,t}$ | | | 0.0114 | 0.0064 |
| | | | (1.1733) | (1.1650) |
| $du_{i,t} \times After1_{i,t}$ | | | 0.0235** | 0.0131** |
| | | | (2.2914) | (2.2649) |
| $du_{i,t} \times After2_{i,t}$ | | | 0.0165 | 0.0108* |
| | | | (1.5909) | (1.8374) |
| $du_{i,t} \times After3_{i,t}$ | | | 0.0233** | 0.0121** |
| | | | (2.2465) | (2.0472) |
| $Growth_{i,t}$ | 0.0000*** | 0.0000*** | 0.0000*** | 0.0000*** |
| | (17.2911) | (20.2095) | (17.7345) | (20.1354) |
| $Roa_{i,t}$ | -0.0116 | -0.0079 | -0.0118 | -0.0080 |
| | (-1.1321) | (-1.2781) | (-1.1524) | (-1.2982) |
| $Lev_{i,t}$ | 0.0002 | -0.0000 | 0.0002 | -0.0000 |
| | (1.1414) | (-0.4227) | (1.2276) | (-0.3685) |
| $Maghold_{i,t}$ | -0.0236 | -0.0134 | -0.0233 | -0.0132 |
| | (-1.5166) | (-1.4156) | (-1.4882) | (-1.3872) |
| $FirmAge_{i,t}$ | 0.1531*** | 0.0990*** | 0.1543*** | 0.0994*** |
| | (3.4942) | (4.0387) | (3.5177) | (4.0545) |
| $Size_{i,t}$ | 0.0507*** | 0.0234*** | 0.0514*** | 0.0237*** |
| | (6.5207) | (5.3970) | (6.6124) | (5.4715) |
| $Tophold_{i,t}$ | -0.1078** | -0.0623** | -0.1101** | -0.0633** |
| | (-2.0009) | (-2.0558) | (-2.0422) | (-2.0887) |
| $Dual_{i,t}$ | -0.0012 | -0.0010 | -0.0009 | -0.0009 |
| | (-0.1286) | (-0.1938) | (-0.1022) | (-0.1705) |
| $Soe_{i,t}$ | 0.0275 | 0.0130 | 0.0279 | 0.0132 |
| | (1.2402) | (1.0966) | (1.2526) | (1.1115) |
| $HHI_{i,t}$ | -0.0153 | 0.0040 | -0.0169 | 0.0035 |
| | (-0.1811) | (0.0879) | (-0.2003) | (0.0780) |
| $CI_{j,t}$ | -0.0229** | -0.0126** | -0.0243** | -0.0132** |
| | (-2.2113) | (-2.4145) | (-2.3430) | (-2.5324) |
| _cons | -0.8943*** | -1.4253*** | -0.9125*** | -1.4326*** |
| | (-5.0552) | (-14.3042) | (-5.1567) | (-14.3983) |
| Firm FE | Yes | Yes | Yes | Yes |
| Year FE | Yes | Yes | Yes | Yes |

| Observations | 27033 | 27033 | 27033 | 27033 |
| --- | --- | --- | --- | --- |
| adj. R2 | 0.047 | 0.042 | 0.046 | 0.042 |

The table reports the estimated results from the difference-in-difference regressions,t statistic based on the robust standard error is in parentheses.***, **, and * indicate significance at the 1%, 5%, and 10% levels, respectively.

## 6| ORGANIZATION COSTS CHANNEL

As mentioned above, the impact of enterprise digital transformation on enterprise diversification through economies from internalizing transactions depends on whether enterprise digital transformation plays a leading role in reducing transaction costs or reducing enterprise organization costs.If digitization plays a dominant role in reducing organization costs,Then, the improvement of enterprise digitization level should be more conducive to those enterprises with high organization costs, which will play a more significant role in promoting the diversification of such enterprises. This paper intends to further explore whether there is heterogeneity in the promotion effect of enterprise digitization on enterprise diversification in enterprises with different organization costs. In order to verify the above speculation, this paper uses the following two indicators to measure the internal control cost of enterprises.

(1)Proportion of administrative expenses. This index directly reflects the organization costs of enterprises. Specifically, according to the sample median of the proportion of enterprise management expenses in operating revenue, we conduct sub sample regression. According to the regression results in Table 6 (1) - (4), we find that the promotion effect of enterprise digitization on enterprise diversification is more significant in the group with higher management expenses, however, the coefficient difference between groups was not statistically significant by the suest test,which cannot verifies the above discussion.

(2) Enterprise growth. According to the enterprise life cycle theory, enterprises in the early stage of growth usually have simple organizational structure and low organization costs; In contrast, when the enterprise is in the mature or declining stage, the organizational structure is complex and the organization costs increases gradually. Therefore, enterprise growth can reflect the organization costs faced by enterprises to a certain extent. The higher the growth, the lower the organization costs. This paper carries out grouping regression according to the median growth rate of operating revenue. According to the regression results in columns (1) - (4) of table 7, the promotion effect of enterprise digitization on enterprise diversification is more significant in the group with low enterprise growth rate, however, the coefficient difference between groups was not statistically significant by the suest test, which cannot verifies the above discussion. Based on the above discussion, hypothesis 1 does not hold.

**Table 6 Digitization on diversification in high management cost group and low management cost group**

| Model | FE | | | |
|---|---|---|---|---|
| | High organization cost | Low organization cost | High organization cost | Low organization cost |
| | High management cost group | low management cost group | High management cost group | low management cost group |
| | (1) | (2) | (3) | (4) |
| Dependent variable | Total entropy | Total entropy | Herfindahl index | Herfindahl index |
| $Digitization_{i,t}$ | 0.0178*** | 0.0139** | 0.0097*** | 0.0075* |
| | (3.0063) | (2.0302) | (2.9204) | (1.9289) |
| $Growth_{i,t}$ | 0.0000*** | 0.0000* | 0.0000*** | 0.0000 |
| | (17.2142) | (1.6882) | (17.8639) | (1.6056) |
| $Roa_{i,t}$ | -0.0053 | -0.0062 | -0.0038 | -0.0098 |
| | (-0.5733) | (-0.2270) | (-0.6768) | (-0.6546) |
| $Lev_{i,t}$ | 0.0001 | -0.0018 | -0.0000 | -0.0049 |
| | (0.6970) | (-0.0898) | (-0.4804) | (-0.4420) |
| $Maghold_{i,t}$ | -0.0181 | -0.0731 | -0.0105 | -0.0416 |
| | (-1.0255) | (-1.0339) | (-1.0170) | (-1.0793) |
| $FirmAge_{i,t}$ | 0.2627*** | 0.0620 | 0.1561*** | 0.0396 |
| | (4.2529) | (1.1094) | (4.4604) | (1.2367) |
| $Size_{i,t}$ | 0.0606*** | 0.0521*** | 0.0330*** | 0.0259*** |
| | (5.7894) | (4.5564) | (5.5549) | (3.9683) |
| $Tophold_{i,t}$ | -0.0696 | -0.1159* | -0.0516 | -0.0603 |
| | (-0.9150) | (-1.7659) | (-1.2046) | (-1.6168) |
| $Dual_{i,t}$ | -0.0007 | 0.0108 | -0.0007 | 0.0044 |
| | (-0.0648) | (0.9185) | (-0.1118) | (0.6653) |
| $Soe_{i,t}$ | 0.0418 | 0.0161 | 0.0177 | 0.0060 |
| | (1.5239) | (0.5230) | (1.1495) | (0.3640) |
| $HHI_{i,t}$ | 0.0285 | 0.0233 | 0.0206 | 0.0096 |
| | (0.2409) | (0.2380) | (0.3126) | (0.1783) |
| $CI_{j,t}$ | -0.0146 | -0.0213 | -0.0097 | -0.0109 |
| | (-0.9477) | (-1.5392) | (-1.0425) | (-1.6225) |
| _cons | -1.3264*** | -0.7937*** | -1.7338*** | -1.3864*** |
| | (-5.3758) | (-2.9704) | (-12.4609) | (-9.1772) |
| Firm FE | Yes | Yes | Yes | Yes |
| Year FE | Yes | Yes | Yes | Yes |
| Observations | 16177 | 16208 | 16177 | 16208 |
| adj. R2 | 0.061 | 0.046 | 0.063 | 0.039 |
| SUEST test | chi2( 1) = 0.19 Prob > chi2 = 0.6593 | | chi2( 1) = 0.19 Prob > chi2 = 0.6657 | |

t statistic based on the robust standard error is in parentheses.***, **, and * indicate significance at the 1%, 5%, and 10% levels, respectively.

**Table 7 Digitization on diversification in high growth rate group and low growth rate group**

| Model | FE | | | |
|---|---|---|---|---|
| | Low organization cost | High organization cost | Low organization cost | High organization cost |
| | High growth rate group | low growth rate group | High growth rate group | low growth rate group |
| | (1) | (2) | (3) | (4) |
| Dependent variable | Total entropy | Total entropy | Herfindahl index | Herfindahl index |
| $Digitization_{i,t}$ | 0.0181*** | 0.0205*** | 0.0092*** | 0.0105*** |
| | (3.3841) | (3.5353) | (3.0750) | (3.2418) |
| $Growth_{i,t}$ | -0.0000 | -0.0677*** | 0.0000 | -0.0414*** |
| | (-0.3653) | (-3.6040) | (1.3470) | (-3.8201) |
| $Roa_{i,t}$ | -0.0244 | -0.0157 | -0.0180 | -0.0132* |
| | (-0.4933) | (-1.3024) | (-0.6109) | (-1.8575) |
| $Lev_{i,t}$ | -0.0015 | 0.0002 | -0.0020 | -0.0000 |
| | (-0.1436) | (1.2958) | (-0.3285) | (-0.4571) |
| $Maghold_{i,t}$ | -0.0174* | -0.1003* | -0.0091 | -0.0684** |
| | (-1.8459) | (-1.7705) | (-1.6328) | (-2.1530) |
| $FirmAge_{i,t}$ | 0.1775*** | 0.1513*** | 0.1183*** | 0.0971*** |
| | (3.6088) | (2.7917) | (4.3085) | (3.2040) |
| $Size_{i,t}$ | 0.0465*** | 0.0554*** | 0.0228*** | 0.0256*** |
| | (5.3275) | (5.9105) | (4.5848) | (4.9146) |
| $Tophold_{i,t}$ | -0.1047* | -0.1405** | -0.0601* | -0.0824** |
| | (-1.7671) | (-2.3051) | (-1.7968) | (-2.4108) |
| $Dual_{i,t}$ | -0.0054 | 0.0056 | -0.0048 | 0.0048 |
| | (-0.4761) | (0.5627) | (-0.7238) | (0.8232) |
| $Soe_{i,t}$ | -0.0116 | 0.0427* | -0.0065 | 0.0204 |
| | (-0.4553) | (1.6869) | (-0.4658) | (1.5001) |
| $HHI_{i,t}$ | -0.0736 | 0.0076 | -0.0269 | 0.0171 |
| | (-0.7624) | (0.0814) | (-0.5064) | (0.3459) |
| $CI_{j,t}$ | -0.0195 | -0.0246* | -0.0153*** | -0.0121* |
| | (-1.5347) | (-1.8211) | (-2.7925) | (-1.7779) |
| _cons | -0.8629*** | -0.9573*** | -1.4548*** | -1.4505*** |
| | (-4.3724) | (-4.3291) | (-12.9679) | (-11.6615) |
| Firm FE | Yes | Yes | Yes | Yes |
| Year FE | Yes | Yes | Yes | Yes |
| Observations | 16476 | 16431 | 16476 | 16431 |
| adj. R2 | 0.047 | 0.057 | 0.044 | 0.052 |
| SUEST | chi2( 1) = | 0.07 | chi2( 1) = | 0.06 |
| test | Prob > chi2 = | 0.7905 | Prob > chi2 = | 0.8002 |

t statistic based on the robust standard error is in parentheses.***, **, and * indicate significance at the 1%, 5%, and 10% levels, respectively.

## 7| TRANSACTION COSTS CHANNEL

Enterprise digital transformation can reduce the transaction costs of enterprises, so whether the impact of enterprise digital transformation on the choice of enterprise diversification strategy is heterogeneous under different degrees of transaction costs. We believe that when the transaction costs of enterprises is low, the space for reducing the transaction costs by digitization is limited. At this time, the effect of digitization on reducing the organization costs will be more obvious, and thus the effect on improving the diversification level of enterprises will be more significant. Because it is difficult to measure enterprise transaction costs directly, this paper measures enterprise transaction costs from enterprise and region levels respectively.

(1) At the enterprise level, this study uses enterprise asset specificity to reflect transaction costs. Enterprises with higher asset specificity face higher lock-in costs because they are often at a disadvantage in the transactions with competitors and face higher transaction costs (Williamson, 2007). In this paper, the proportion of intangible assets in total assets is used to measure the asset specificity of enterprises (Collis and Montgomery, 1997). According to the quartile of the proportion of intangible assets in total assets, sub sample regression is carried out. According to the regression results in Table 8 (1) - (4), we find that the promotion of enterprise digitization on the level of enterprise diversification is more significant in the group with relatively low proportion of intangible assets, however, the coefficient difference between groups was not statistically significant by the suest test,which cannot verifies the above discussion.

(2) At the regional level, this paper studies the marketization level to reflect the transaction costs. In areas with low marketization level, the trading environment is poor, the probability of default among market subjects is higher, and the transaction costs faced by enterprises is often higher. This paper uses the marketization index of the province where the enterprise is located to measure the median of the regional marketization level for sub sample regression. According to the regression results in Table 9 (1) - (4), we find that the promotion effect of enterprise digitization on the level of enterprise diversification is more significant in the group with higher marketization degree, however, the coefficient difference between groups was not statistically significant by the suest test,which cannot verifies the above discussion. Based on the above discussion, hypothesis 2 does not hold.

**Table 8 Digitization on diversification in high intangible asset ratio group and low intangible asset ratio group**

| Model | FE | | | |
|---|---|---|---|---|
| | High transaction costs | Low transaction costs | High transaction costs | Low transaction costs |
| | high intangible asset ratio group | low intangible asset ratio group | high intangible asset ratio group | low intangible asset ratio group |
| Dependent variable | Total entropy | Total entropy | Herfindahl index | Herfindahl index |
| | (1) | (2) | (3) | (4) |
| $Digitization_{i,t}$ | 0.0163** | 0.0243*** | 0.0102** | 0.0133*** |
| | (2.0548) | (3.0981) | (2.1749) | (3.0427) |
| $Growth_{i,t}$ | 0.0000*** | 0.0000 | 0.0000*** | 0.0000 |
| | (7.4326) | (0.7766) | (9.4542) | (0.3084) |
| $Roa_{i,t}$ | -0.0310 | 0.0019 | -0.0181 | -0.0003 |
| | (-1.2702) | (0.0886) | (-1.1321) | (-0.0265) |
| $Lev_{i,t}$ | -0.0003 | 0.0000 | 0.0001 | -0.0001 |
| | (-0.0245) | (0.0094) | (0.0184) | (-0.8382) |
| $Maghold_{i,t}$ | -0.1297* | -0.0232 | -0.0866* | -0.0242 |
| | (-1.7067) | (-0.3638) | (-1.9500) | (-0.6514) |
| $FirmAge_{i,t}$ | 0.0516 | 0.1080 | 0.0278 | 0.0861* |
| | (0.5895) | (1.1594) | (0.5630) | (1.7479) |
| $Size_{i,t}$ | 0.0648*** | 0.0077 | 0.0358*** | -0.0017 |
| | (4.0956) | (0.7174) | (4.1118) | (-0.2815) |
| $Tophold_{i,t}$ | 0.0141 | -0.0463 | -0.0029 | -0.0091 |
| | (0.1425) | (-0.5237) | (-0.0509) | (-0.1895) |
| $Dual_{i,t}$ | -0.0266* | 0.0088 | -0.0111 | 0.0043 |
| | (-1.7083) | (0.5814) | (-1.2073) | (0.5169) |
| $Soe_{i,t}$ | 0.0982*** | -0.0191 | 0.0447** | -0.0172 |
| | (2.6847) | (-0.4456) | (2.2717) | (-0.7754) |
| $HHI_{i,t}$ | -0.2091 | -0.1434 | -0.1545* | -0.0505 |
| | (-1.4006) | (-0.8149) | (-1.8198) | (-0.5512) |
| $CI_{j,t}$ | -0.0347*** | 0.4063*** | -0.0175*** | 0.2073** |
| | (-7.1379) | (2.6690) | (-6.5033) | (2.5003) |
| _cons | -1.0043*** | 0.0898 | -1.5335*** | -0.8817*** |
| | (-2.7772) | (0.2998) | (-7.4183) | (-5.4062) |
| Firm FE | Yes | Yes | Yes | Yes |
| Year FE | Yes | Yes | Yes | Yes |
| Observations | 8239 | 8236 | 8239 | 8236 |
| adj. R2 | 0.065 | 0.025 | 0.064 | 0.016 |
| SUEST test | chi2( 1) = 0.37 Prob > chi2 = 0.5437 | | chi2( 1) = 0.01 Prob > chi2 = 0.9263 | |

t statistic based on the robust standard error is in parentheses.***, **, and * indicate significance at the 1%, 5%, and 10% levels, respectively.

**Table 9 Digitization on diversification in high regional marketization degree group and low regional marketization degree group**

| Model | FE | | | |
|---|---|---|---|---|
| | Low transaction costs | High transaction costs | Low transaction costs | High transaction costs |
| | high regional marketization degree group | low regional marketization degree group | high regional marketization degree group | low regional marketization degree group |
| | (1) | (2) | (3) | (4) |
| Dependent variable | Total entropy | Total entropy | Herfindahl index | Herfindahl index |
| $Digitization_{i,t}$ | 0.0213*** | 0.0163** | 0.0111*** | 0.0077* |
| | (3.3609) | (2.1982) | (3.1526) | (1.8178) |
| $Growth_{i,t}$ | 0.0001 | 0.0000*** | 0.0000 | 0.0000*** |
| | (0.8735) | (14.7872) | (0.6196) | (15.0794) |
| $Roa_{i,t}$ | -0.0615* | -0.0110 | -0.0370* | -0.0076 |
| | (-1.7390) | (-1.0218) | (-1.7883) | (-1.1915) |
| $Lev_{i,t}$ | -0.0064 | 0.0004*** | -0.0052 | 0.0001 |
| | (-0.9408) | (2.7893) | (-1.2925) | (1.6381) |
| $Maghold_{i,t}$ | -0.0523 | -0.0212 | -0.0320 | -0.0118 |
| | (-0.9019) | (-1.5121) | (-0.9902) | (-1.3842) |
| $FirmAge_{i,t}$ | 0.2144*** | 0.0361 | 0.1324*** | 0.0438 |
| | (3.8137) | (0.5239) | (4.2334) | (1.1251) |
| $Size_{i,t}$ | 0.0343*** | 0.0649*** | 0.0160** | 0.0309*** |
| | (2.6257) | (6.2984) | (2.2063) | (5.3040) |
| $Tophold_{i,t}$ | -0.1302 | -0.1263* | -0.0831* | -0.0616 |
| | (-1.5830) | (-1.8435) | (-1.8211) | (-1.5589) |
| $Dual_{i,t}$ | 0.0012 | -0.0065 | 0.0005 | -0.0030 |
| | (0.0919) | (-0.5030) | (0.0615) | (-0.4045) |
| $Soe_{i,t}$ | -0.0016 | 0.0430 | -0.0065 | 0.0219 |
| | (-0.0374) | (1.6124) | (-0.3007) | (1.4869) |
| $HHI_{i,t}$ | 0.0050 | -0.1091 | 0.0085 | -0.0374 |
| | (0.0397) | (-1.1035) | (0.1234) | (-0.6834) |
| $CI_{j,t}$ | 0.0260 | -0.0829 | 0.0127 | -0.0874 |
| | (0.1921) | (-0.3786) | (0.1683) | (-0.7251) |
| _cons | -0.6712** | -0.9314*** | -1.3299*** | -1.4645*** |
| | (-2.3348) | (-3.7996) | (-8.3678) | (-10.4098) |
| Firm FE | Yes | Yes | Yes | Yes |
| Year FE | Yes | Yes | Yes | Yes |
| Observations | 13641 | 15967 | 13641 | 15967 |
| adj. R2 | 0.058 | 0.050 | 0.059 | 0.039 |
| SUEST test | chi2(1) = 0.28 Prob > chi2 = 0.5985 | | chi2(1) = 0.41 Prob > chi2 = 0.5226 | |

t statistic based on the robust standard error is in parentheses.***, **, and * indicate significance at the 1%, 5%, and 10% levels, respectively.

## 8| THE IMPACT OF THE INDUSTRY TYPE

Manufacturing and services. On the one hand, the impact of digitization on the level of enterprise diversification is affected by the degree of transformation. In the early stage of digital transformation, digital technology has not been integrated with enterprise production and operation, so the impact of digitization on enterprise diversification is difficult to be reflected; With the in-depth development of digital transformation, digital technology will be gradually integrated with enterprise production and operation, and the impact of digitization on enterprise organization cost will be gradually reflected. On the whole, the digital transformation process of service industry precedes that of manufacturing industry. Therefore, the role of digitization in promoting the diversification level of enterprises may be more significant in service enterprises. According to the regression results in table 10 (1) - (4), we find that the promotion effect of enterprise digital transformation on enterprise diversification is more significant in the service industry, however, the coefficient difference between groups was not statistically significant by the suest test, which cannot verifies the above discussion.

**Table 10 Digitization on diversification in Manufacturing and Service industry**

| Model | FE | | | |
|---|---|---|---|---|
| | (1) | (2) | (3) | (4) |
| | Service industry | Manufacturing | Service industry | Manufacturing |
| Dependent variable | Total entropy | Total entropy | Herfindahl index | Herfindahl index |
| $Digitization_{i,t}$ | 0.0263*** | 0.0133** | 0.0138*** | 0.0076** |
| | (3.1710) | (2.4663) | (2.9758) | (2.4611) |
| $Growth_{i,t}$ | 0.0000*** | -0.0002** | 0.0000*** | -0.0001** |
| | (10.7885) | (-2.4651) | (14.0006) | (-2.5313) |
| $Roa_{i,t}$ | 0.0219 | -0.0615*** | 0.0064 | -0.0354*** |
| | (0.9019) | (-3.5320) | (0.4467) | (-3.7447) |
| $Lev_{i,t}$ | 0.0005 | -0.0088** | -0.0002 | -0.0057** |
| | (0.3184) | (-2.1327) | (-0.2344) | (-2.3569) |
| $Maghold_{i,t}$ | -0.0094* | -0.1156** | -0.0042 | -0.0724** |
| | (-1.7405) | (-2.2039) | (-1.3462) | (-2.4568) |
| $FirmAge_{i,t}$ | 0.1692** | 0.1226** | 0.1067** | 0.0834*** |
| | (1.9744) | (2.2891) | (2.2784) | (2.7285) |
| $Size_{i,t}$ | 0.0321*** | 0.0607*** | 0.0112* | 0.0316*** |
| | (2.6216) | (5.8598) | (1.6721) | (5.3890) |
| $Tophold_{i,t}$ | -0.0183 | -0.2045*** | -0.0022 | -0.1184*** |
| | (-0.1926) | (-3.0272) | (-0.0422) | (-3.0992) |
| $Dual_{i,t}$ | -0.0033 | 0.0042 | -0.0021 | 0.0021 |
| | (-0.2033) | (0.4084) | (-0.2298) | (0.3496) |
| $Soe_{i,t}$ | 0.0230 | 0.0266 | 0.0091 | 0.0145 |
| | (0.6089) | (0.9919) | (0.4536) | (0.9913) |
| $HHI_{i,t}$ | -0.0436 | -0.1397 | -0.0010 | -0.0627 |

|  | (-0.3069) | (-1.3489) | (-0.0127) | (-1.0675) |
| --- | --- | --- | --- | --- |
| $CI_{j,t}$ | -0.0218** | 4.4860** | -0.0114** | 2.2333** |
|  | (-2.0275) | (2.4812) | (-2.1433) | (2.2226) |
| _cons | -0.5292* | -1.0141*** | -1.1906*** | -1.5517*** |
|  | (-1.7321) | (-4.3142) | (-7.1322) | (-11.6420) |
| Firm FE | Yes | Yes | Yes | Yes |
| Year FE | Yes | Yes | Yes | Yes |
| Observations | 9750 | 19554 | 9750 | 19554 |
| adj. R2 | 0.046 | 0.063 | 0.037 | 0.061 |
| SUEST | chi2( 1) = | 2.00 | chi2( 1) = | 1.48 |
| test | Prob > chi2 = | 0.1571 | Prob > chi2 = | 0.2238 |

t statistic based on the robust standard error is in parentheses.***, **, and * indicate significance at the 1%, 5%, and 10% levels, respectively.

## 9| MARKET POWER CHANNEL

We use the HHI index to measure industry concentration ($HHI2_{i,t}$ The square sum of the companyi's Total asset in the proportion of all companies in the industry in year t). We performed grouping regression according to the annual median of industry concentration, according to the regression results in Table 11 (1) - (2), we find that the impact of enterprise digital transformation on enterprise diversification is more significant in the group with high market concentration, and the coefficient difference between groups passed the suest test. It shows that the pursuit of market monopoly power and interests is an important reason for the diversification level of enterprises caused by enterprise digital transformation.

To ensure reliable results, we use the annual median of $HHI_{i,t}$ (The square sum of the companyi's operating income in the proportion of all companies in the industry in year t) as the grouping variable for regression, according to Table 11(3)-(4) The regression results of , the coefficient difference between groups passed the suest test, further illustrate the robustness of the benchmark conclusions. Based on the above discussion, hypothesis 3 holds.

**Table 11 Digitization on diversification in High market concentration group and Low market concentration group**

| Model | FE | | | |
|---|---|---|---|---|
| | $HHI2_{i,t}$ | | $HHI_{i,t}$ | |
| | High market concentration group (1) | Low market concentration group (2) | High market concentration group (3) | Low market concentration group (4) |
| Dependent variable | Total entropy | Total entropy | Total entropy | Total entropy |
| $Digitization_{i,t}$ | 0.0306*** | 0.0123* | 0.0270*** | 0.0117* |
| | (4.9386) | (1.9238) | (4.3142) | (1.9052) |
| $Growth_{i,t}$ | 0.0000*** | 0.0000 | 0.0000*** | -0.0000 |
| | (17.4014) | (0.7323) | (16.9140) | (-0.4356) |
| $Roa_{i,t}$ | -0.0020 | -0.0834** | 0.0029 | -0.0821* |
| | (-0.2536) | (-2.2808) | (0.3333) | (-1.8625) |
| $Lev_{i,t}$ | 0.0003** | 0.0018 | 0.0003* | 0.0076 |
| | (2.0271) | (0.2165) | (1.7353) | (0.8877) |
| $Maghold_{i,t}$ | -0.0194* | -0.0343 | -0.0187* | -0.0709 |
| | (-1.7216) | (-0.7605) | (-1.7658) | (-1.1869) |
| $FirmAge_{i,t}$ | 0.1631*** | 0.1428** | 0.2177*** | 0.0627 |
| | (2.8157) | (2.2203) | (3.7351) | (1.0147) |
| $Size_{i,t}$ | 0.0592*** | 0.0463*** | 0.0534*** | 0.0493*** |
| | (5.5448) | (4.4367) | (5.0370) | (4.6333) |
| $Tophold_{i,t}$ | -0.1157 | -0.2231*** | -0.0664 | -0.2554*** |
| | (-1.6405) | (-2.8349) | (-0.9396) | (-3.3026) |
| $Dual_{i,t}$ | -0.0018 | 0.0100 | -0.0054 | 0.0088 |
| | (-0.1501) | (0.8636) | (-0.4565) | (0.7662) |
| $Soe_{i,t}$ | 0.0408 | 0.0158 | 0.0356 | 0.0284 |
| | (1.4217) | (0.5164) | (1.2767) | (0.9122) |
| $HHI_{i,t}$ | 0.0300 | -0.0112 | 0.0208 | -0.0437 |
| | (0.3869) | (-0.0302) | (0.2631) | (-0.0884) |
| $CI_{j,t}$ | -0.0239** | 4.5160*** | -0.0189* | -1.5900** |
| | (-2.3374) | (2.7473) | (-1.8509) | (-2.2074) |
| _cons | -1.1346*** | -0.7472*** | -1.1290*** | -0.6094** |
| | (-4.6755) | (-3.0520) | (-4.6369) | (-2.4691) |
| Firm FE | Yes | Yes | Yes | Yes |
| Year FE | Yes | Yes | Yes | Yes |
| Observations | 16145 | 16762 | 16062 | 16845 |
| adj. R2 | 0.061 | 0.049 | 0.049 | 0.054 |
| SUEST test | chi2(1) = 4.98 Prob > chi2 = 0.0256 | | chi2(1) = 4.29 Prob > chi2 = 0.0384 | |

t statistic based on the robust standard error is in parentheses.***, **, and * indicate significance at the 1%, 5%, and 10% levels, respectively.

## 10| BLOCKHOLDER CONTROL CHANNEL

To test whether digital transformation increases diversification through block holders' intervention, Following Gu et al. (2018), we proxy block holder ownership, we add a direct proxy for block holders' ownership, $BlockHolding_{i,t}$, which is the total holding of block holders. We conducted group regression according to the annual median block holders shareholding. According to the regression results in columns (1) - (4) of table 12, the difference of correlation coefficient between groups did not pass the suest test, indicating that this channel hypothesis is not tenable. Based on the above discussion, Hypothesis 4 does not hold.

**Table 12 Digitization on diversification in High block holders shareholding group and Low block holders shareholding group**

| Model | FE | | | |
|---|---|---|---|---|
| | High block holders shareholding group | Low block holders shareholding group | High block holders shareholding group | Low block holders shareholding group |
| | (1) | (2) | (3) | (4) |
| Dependent variable | Total entropy | Total entropy | Herfindahl index | Herfindahl index |
| $Digitization_{i,t}$ | 0.0175*** | 0.0193*** | 0.0100*** | 0.0085*** |
| | (2.7564) | (3.2449) | (2.7381) | (2.5978) |
| $Growth_{i,t}$ | 0.0000** | 0.0000*** | 0.0000 | 0.0000*** |
| | (2.2601) | (15.4420) | (1.5664) | (17.5738) |
| $Roa_{i,t}$ | -0.0404 | -0.0108 | -0.0231 | -0.0085 |
| | (-1.3488) | (-1.2305) | (-1.4129) | (-1.5874) |
| $Lev_{i,t}$ | 0.0335* | 0.0003** | 0.0175 | 0.0000 |
| | (1.7189) | (2.0184) | (1.4773) | (0.5798) |
| $Maghold_{i,t}$ | -0.0161* | 0.0461 | -0.0086 | 0.0335 |
| | (-1.6972) | (0.7023) | (-1.4603) | (0.9320) |
| $FirmAge_{i,t}$ | 0.0961 | 0.2054*** | 0.0721** | 0.1278*** |
| | (1.6306) | (3.1665) | (2.1148) | (3.5795) |
| $Size_{i,t}$ | 0.0458*** | 0.0628*** | 0.0231*** | 0.0311*** |
| | (3.9179) | (6.2189) | (3.3475) | (5.6658) |
| $Tophold_{i,t}$ | -0.2938*** | -0.0804 | -0.1863*** | -0.0374 |
| | (-3.1838) | (-1.0667) | (-3.5358) | (-0.9093) |
| $Dual_{i,t}$ | 0.0042 | -0.0047 | 0.0003 | -0.0022 |
| | (0.3363) | (-0.4384) | (0.0455) | (-0.3642) |
| $Soe_{i,t}$ | -0.0133 | 0.0710** | -0.0081 | 0.0363** |
| | (-0.4927) | (2.1161) | (-0.5210) | (2.0595) |
| $HHI_{i,t}$ | -0.0008 | -0.0738 | 0.0201 | -0.0432 |
| | (-0.0062) | (-0.7534) | (0.2970) | (-0.8335) |
| $CI_{j,t}$ | -0.0361*** | -0.0164 | -0.0167*** | -0.0101* |
| | (-4.5874) | (-1.5947) | (-3.7062) | (-1.7271) |
| _cons | -0.6251** | -1.2917*** | -1.3342*** | -1.6693*** |
| | (-2.3964) | (-5.5165) | (-8.7545) | (-12.9310) |
| Firm FE | Yes | Yes | Yes | Yes |
| Year FE | Yes | Yes | Yes | Yes |
| Observations | 13936 | 18971 | 13936 | 18971 |
| adj. R2 | 0.054 | 0.047 | 0.053 | 0.041 |
| SUEST test | chi2( 1) = | 0.04 | chi2( 1) = | 0.09 |
| | Prob > chi2 = | 0.8399 | Prob > chi2 = | 0.7661 |

t statistic based on the robust standard error is in parentheses.***, **, and * indicate significance at the 1%, 5%, and 10% levels, respectively.

## 11| INFORMATION ASYMMETRY CHANNEL

The presence of external monitoring systems, including analysts' coverage and external auditors, is attributable to a more transparent information environment for outsiders (see Choi & Lee, 2013; Zuckerman, 2000).

To verify whether information asymmetry the key channel affecting the main effect of this paper, we follow the literature and use two proxies for information asymmetry that have been widely adopted: the number of analysts following a firm ($Analyst_{i,t}$); the number of analysts' reports covering a firm ($Report_{i,t}$).

Firstly, we conducted grouping regression according to the annual median concerned by analyst focus. According to the regression results in table 13 (1) - (4), the difference of regression coefficient between groups failed to pass the suest test. Then, we conducted grouping regression according to the annual median of interest in the research report. According to the regression results in table 14 (1) - (4), the difference of regression coefficient between groups failed to pass the suest test. To sum up, this channel assumption is not tenable. Based on the above discussion, Hypothesis 5 does not hold.

**Table 13 Digitization on diversification in High analyst focus group and Low analyst focus group**

| Model | FE | | | |
|---|---|---|---|---|
| | High analyst focus group | Low analyst focus group | High analyst focus group | Low analyst focus group |
| | (1) | (2) | (3) | (4) |
| Dependent variable | Total entropy | Total entropy | Herfindahl index | Herfindahl index |
| $Digitization_{i,t}$ | 0.0160*** | 0.0219*** | 0.0072** | 0.0130*** |
| | (2.9054) | (3.5003) | (2.4776) | (3.6374) |
| $Growth_{i,t}$ | 0.0000*** | 0.0000*** | 0.0000*** | 0.0000*** |
| | (2.7876) | (19.3594) | (3.6069) | (19.8761) |
| $Roa_{i,t}$ | -0.0937 | -0.0067 | -0.0481 | -0.0057 |
| | (-1.4774) | (-0.8675) | (-1.3273) | (-1.2012) |
| $Lev_{i,t}$ | 0.1035** | 0.0001 | 0.0564** | -0.0000 |
| | (2.3152) | (0.9177) | (2.2671) | (-0.4766) |
| $Maghold_{i,t}$ | -0.0129* | -0.0194 | -0.0073 | -0.0148 |
| | (-1.7781) | (-0.5083) | (-1.5938) | (-0.6691) |
| $FirmAge_{i,t}$ | 0.0913* | 0.1687*** | 0.0660** | 0.1044*** |
| | (1.6664) | (2.7652) | (2.1672) | (3.0378) |
| $Size_{i,t}$ | 0.0641*** | 0.0494*** | 0.0359*** | 0.0238*** |
| | (5.0416) | (5.3340) | (5.0231) | (4.5628) |
| $Tophold_{i,t}$ | -0.1087 | -0.1138* | -0.0636 | -0.0659* |
| | (-1.3697) | (-1.7597) | (-1.4748) | (-1.7703) |
| $Dual_{i,t}$ | -0.0016 | 0.0030 | -0.0021 | 0.0019 |
| | (-0.1326) | (0.3023) | (-0.2972) | (0.3150) |
| $Soe_{i,t}$ | -0.0209 | 0.0093 | -0.0075 | 0.0017 |
| | (-0.6261) | (0.3963) | (-0.4137) | (0.1334) |
| $HHI_{i,t}$ | -0.1815 | 0.0515 | -0.0627 | 0.0298 |
| | (-1.5791) | (0.5285) | (-1.0207) | (0.5666) |
| $CI_{j,t}$ | -0.0213 | -0.0265** | -0.0102 | -0.0159** |
| | (-1.4173) | (-2.0766) | (-1.4632) | (-1.9767) |
| _cons | -1.1546*** | -0.8726*** | -1.6833*** | -1.4288*** |
| | (-3.9588) | (-3.9225) | (-10.2465) | (-11.3284) |
| Firm FE | Yes | Yes | Yes | Yes |
| Year FE | Yes | Yes | Yes | Yes |
| Observations | 15534 | 17373 | 15534 | 17373 |
| adj. R2 | 0.068 | 0.042 | 0.065 | 0.039 |
| SUEST test | chi2( 1) = 0.43 Prob > chi2 = 0.5121 | | chi2( 1) = 1.36 Prob > chi2 = 0.2429 | |

t statistic based on the robust standard error is in parentheses.***, **, and * indicate significance at the 1%, 5%, and 10% levels, respectively.

**Table 14 Digitization on diversification in High research report Concern Group and Low Research Report Concern Group**

| Model | FE | | | |
|---|---|---|---|---|
| | High research report Concern Group | Low Research Report Concern Group | High research report Concern Group | Low Research Report Concern Group |
| | (1) | (2) | (3) | (4) |
| Dependent variable | Total entropy | Total entropy | Herfindahl index | Herfindahl index |
| $Digitization_{i,t}$ | 0.0167*** | 0.0205*** | 0.0077*** | 0.0122*** |
| | (3.0135) | (3.2754) | (2.6415) | (3.3861) |
| $Growth_{i,t}$ | 0.0000*** | 0.0000*** | 0.0000*** | 0.0000*** |
| | (2.5902) | (19.0926) | (3.3581) | (19.8156) |
| $Roa_{i,t}$ | -0.1263* | -0.0064 | -0.0663* | -0.0054 |
| | (-1.9247) | (-0.8197) | (-1.7919) | (-1.1508) |
| $Lev_{i,t}$ | 0.0853* | 0.0001 | 0.0493* | -0.0000 |
| | (1.8904) | (1.1322) | (1.9543) | (-0.3079) |
| $Maghold_{i,t}$ | -0.0119* | -0.0190 | -0.0067 | -0.0151 |
| | (-1.7213) | (-0.5116) | (-1.5521) | (-0.6966) |
| $FirmAge_{i,t}$ | 0.1110** | 0.1567** | 0.0749** | 0.0984*** |
| | (1.9990) | (2.5381) | (2.4402) | (2.8419) |
| $Size_{i,t}$ | 0.0598*** | 0.0520*** | 0.0332*** | 0.0251*** |
| | (4.6455) | (5.5540) | (4.5709) | (4.7480) |
| $Tophold_{i,t}$ | -0.0773 | -0.1231* | -0.0487 | -0.0698* |
| | (-0.9579) | (-1.9180) | (-1.1218) | (-1.8866) |
| $Dual_{i,t}$ | 0.0022 | 0.0030 | 0.0001 | 0.0023 |
| | (0.1811) | (0.2949) | (0.0208) | (0.3896) |
| $Soe_{i,t}$ | -0.0292 | 0.0086 | -0.0136 | 0.0017 |
| | (-0.8925) | (0.3695) | (-0.7639) | (0.1331) |
| $HHI_{i,t}$ | -0.1493 | 0.0191 | -0.0461 | 0.0141 |
| | (-1.2840) | (0.2053) | (-0.7421) | (0.2784) |
| $CI_{j,t}$ | -0.0212 | -0.0283** | -0.0103 | -0.0168** |
| | (-1.4362) | (-2.2426) | (-1.5048) | (-2.1404) |
| _cons | -1.1031*** | -0.8925*** | -1.6409*** | -1.4394*** |
| | (-3.6901) | (-3.9532) | (-9.7907) | (-11.2427) |
| Firm FE | Yes | Yes | Yes | Yes |
| Year FE | Yes | Yes | Yes | Yes |
| Observations | 15657 | 17250 | 15657 | 17250 |
| adj. R2 | 0.063 | 0.044 | 0.060 | 0.040 |
| SUEST | chi2( 1) = | 0.19 | chi2( 1) = | 0.82 |
| test | Prob > chi2 = | 0.6623 | Prob > chi2 = | 0.3665 |

t statistic based on the robust standard error is in parentheses.***, **, and * indicate significance at the 1%, 5%, and 10% levels, respectively.

## 12| FIRM RISK CHANNEL

We take every three years as an observation period to calculate the standard deviation ($risk1_{i,t}$) and range ($risk2_{i,t}$) of the industry adjusted ROA respectively. When the enterprise risk is higher than the median of the sample, the value of the dummy variable $H-risk-1(2)_{i,t}$ is 1, otherwise it is 0.we include the $H-risk-1(2)_{i,t}$ and the interaction terms of standardized digital transformation and standardized $H-risk-1(2)_{i,t}$ proxies into the baseline regression model.According to the regression results in columns (1) - (2) of table 14, we find that the regression coefficient of the $\overline{Digitization_{i,t}} * \overline{H-risk-1_{i,t}}$ is significantly positive, indicating that the digital transformation of enterprises in high-risk group can better promote the development of enterprise diversification. The regression results in table 15 are consistent with the benchmark regression, which further illustrates the robustness of the conclusion. In addition, we use enterprise risk as the outcome variable to analyze the relationship between enterprise digital transformation and enterprise risk. According to the regression results in columns (1) - (2) of table 16, we find that there is a significant positive correlation between enterprise digital transformation and enterprise risk. Based on the above analysis results, we believe that the digital transformation of enterprises is often accompanied by high firm risks. In order to reduce firm risks, firm will adopt diversified development strategy. This channel assumption is established. Based on the above discussion, hypothesis 6 holds.

**Table 14 Digitization on diversification considers the firm risk**

|  | $risk1_{i,t}$ | |
| --- | --- | --- |
|  | (1) | (2) |
| *Dependent variable* | Total entropy | Herfindahl index |
| $Digitization_{i,t}$ | 0.0168*** | 0.0092*** |
|  | (3.4323) | (3.3038) |
| $\overline{Digitization_{i,t}} * \overline{H-risk-1_{i,t}}$ | 0.0054** | 0.0028** |
|  | (2.4723) | (2.2637) |
| $H-risk-1_{i,t}$ | 0.0022 | 0.0010 |
|  | (0.4661) | (0.3590) |
| $Growth_{i,t}$ | 0.0000*** | 0.0000*** |
|  | (17.0733) | (19.2448) |
| $Roa_{i,t}$ | -0.0117 | -0.0077 |
|  | (-1.1591) | (-1.2985) |
| $Lev_{i,t}$ | 0.0001 | -0.0000 |
|  | (0.9125) | (-0.4626) |
| $Maghold_{i,t}$ | -0.0177 | -0.0098 |
|  | (-1.5848) | (-1.4631) |
| $FirmAge_{i,t}$ | 0.1703*** | 0.1064*** |
|  | (3.3343) | (3.6947) |
| $Size_{i,t}$ | 0.0434*** | 0.0211*** |
|  | (5.5969) | (4.7957) |
| $Tophold_{i,t}$ | -0.1013* | -0.0597** |
|  | (-1.8873) | (-1.9672) |
| $Dual_{i,t}$ | -0.0027 | -0.0026 |

|  | (-0.3069) | (-0.5034) |
| --- | --- | --- |
| $Soe_{i,t}$ | 0.0370* | 0.0163 |
|  | (1.7414) | (1.4105) |
| $HHI_{i,t}$ | 0.0231 | 0.0033 |
|  | (0.2660) | (0.0694) |
| $CI_{j,t}$ | -0.0219** | -0.0123** |
|  | (-2.1477) | (-2.3912) |
| _cons | -0.7993*** | -1.3982*** |
|  | (-4.2405) | (-13.2066) |
| Firm FE | Yes | Yes |
| Year FE | Yes | Yes |
| Observations | 29269 | 29269 |
| adj. $R^2$ | 0.041 | 0.038 |

t statistic based on the robust standard error is in parentheses.***, **, and * indicate significance at the 1%, 5%, and 10% levels, respectively.

**Table 15 Digitization on diversification considers the firm risk**

|  | $risk2_{i,t}$ | |
| --- | --- | --- |
|  | (1) | (2) |
| Dependent variable | Total entropy | Herfindahl index |
| $Digitization_{i,t}$ | 0.0167*** | 0.0092*** |
|  | (3.4254) | (3.2974) |
| $\overline{Digitization_{i,t}} * \overline{H-risk-2_{i,t}}$ | 0.0054** | 0.0027** |
|  | (2.4587) | (2.1685) |
| $H-risk-2_{i,t}$ | 0.0052 | 0.0026 |
|  | (1.1160) | (0.9847) |
| $Growth_{i,t}$ | 0.0000*** | 0.0000*** |
|  | (17.0459) | (19.1734) |
| $Roa_{i,t}$ | -0.0113 | -0.0075 |
|  | (-1.1267) | (-1.2706) |
| $Lev_{i,t}$ | 0.0001 | -0.0000 |
|  | (0.9432) | (-0.4341) |
| $Maghold_{i,t}$ | -0.0176 | -0.0097 |
|  | (-1.5902) | (-1.4677) |
| $FirmAge_{i,t}$ | 0.1699*** | 0.1062*** |
|  | (3.3283) | (3.6888) |
| $Size_{i,t}$ | 0.0437*** | 0.0212*** |
|  | (5.6286) | (4.8260) |
| $Tophold_{i,t}$ | -0.1009* | -0.0595* |
|  | (-1.8806) | (-1.9606) |
| $Dual_{i,t}$ | -0.0027 | -0.0026 |
|  | (-0.3099) | (-0.5061) |
| $Soe_{i,t}$ | 0.0371* | 0.0163 |
|  | (1.7445) | (1.4138) |
| $HHI_{i,t}$ | 0.0225 | 0.0030 |
|  | (0.2596) | (0.0633) |
| $CI_{j,t}$ | -0.0218** | -0.0123** |
|  | (-2.1370) | (-2.3753) |
| _cons | -0.8054*** | -1.4016*** |
|  | (-4.2723) | (-13.2358) |
| Firm FE | Yes | Yes |
| Year FE | Yes | Yes |
| Observations | 29269 | 29269 |
| adj. $R^2$ | 0.041 | 0.038 |

t statistic based on the robust standard error is in parentheses.***, **, and * indicate significance at the 1%, 5%, and 10% levels, respectively.

**Table 16 Digitization on diversification considers the firm risk**

|  | (1) | (2) |
|---|---|---|
| Dependent variable | $risk1_{i,t}$ | $risk2_{i,t}$ |
| $Digitization_{i,t}$ | 0.0023*** | 0.0042*** |
|  | (3.8185) | (3.7883) |
| $Growth_{i,t}$ | 0.0000*** | 0.0000*** |
|  | (7.0737) | (7.6431) |
| $Roa_{i,t}$ | -0.0164** | -0.0298** |
|  | (-2.1636) | (-2.1984) |
| $Lev_{i,t}$ | 0.0001 | 0.0001 |
|  | (0.5171) | (0.5333) |
| $Maghold_{i,t}$ | -0.0036 | -0.0065 |
|  | (-0.7318) | (-0.6967) |
| $FirmAge_{i,t}$ | 0.0397*** | 0.0750*** |
|  | (5.6464) | (5.6588) |
| $Size_{i,t}$ | -0.0187*** | -0.0355*** |

|  | (-14.8347) | (-14.7968) |
|---|---|---|
| $Tophold_{i,t}$ | -0.0255*** | -0.0471*** |
|  | (-3.6589) | (-3.5598) |
| $Dual_{i,t}$ | 0.0015 | 0.0027 |
|  | (0.9374) | (0.9086) |
| $Soe_{i,t}$ | -0.0026 | -0.0050 |
|  | (-0.9719) | (-1.0022) |
| $HHI_{i,t}$ | 0.0383*** | 0.0751*** |
|  | (3.2656) | (3.3616) |
| $CI_{j,t}$ | -0.0051** | -0.0097** |
|  | (-2.0352) | (-2.0378) |
| _cons | 0.3477*** | 0.6602*** |
|  | (12.8793) | (12.8950) |
| Firm FE | Yes | Yes |
| Year FE | Yes | Yes |
| Observations | 30891 | 30891 |
| adj. $R^2$ | 0.139 | 0.140 |

t statistic based on the robust standard error is in parentheses.***, **, and * indicate significance at the 1%, 5%, and 10% levels, respectively.

## 13| DISCUSSIONS

For hypothesis 1 and hypothesis 2 are not true, transaction costs and organizational costs are usually important factors to explain the boundaries of enterprises (Coase, 1937). However, we find that the groups with obvious differences in organizational costs and transaction costs in the digital transformation of enterprises have no significant difference in the impact on the enterprise's diversification. This may be because the boundaries of enterprise diversification strategies are different from the boundaries of enterprises, The boundary of the enterprise's diversification strategy needs to consider the trade-off between the benefits brought by the diversification strategy and the organizational costs brought by the diversification strategy (Jones and hill, 1988). In other words, the digital transformation of the enterprise may have more impact on the benefits side of the diversification strategy. This study shows that the reasons for enterprises to adopt diversification strategy under the digital background are not driven by the traditional reasons for determining enterprise boundaries. This may provide a basis for explaining the controversial reasons for the impact of digital technology on enterprise boundaries, that is, the trend of separation between the discussion of enterprise boundaries and the choice of enterprise diversification strategy under the digital background, That is, the strategic choices of enterprises under different circumstances are based on the actual situation of the enterprise, and these choices may be inefficient, that is, whether the diversification strategy is implemented or not is not considered only from the perspective of transaction costs and organizational costs. This explains why there are two contradictory viewpoints of narrowing the enterprise boundary (lajili and Mahoney, 2006) and improving the enterprise boundary (Luo, 2021) with digital technology, That is, the impact of digital technology on enterprise boundaries may still be in an disequilibrium and unstable state, the market has not reached equilibrium, and it may require further evolution in time and space to give a more clear answer.

    For the establishment of hypothesis 3, there are significant differences in the groups with high and low levels of competition in the digital transformation of enterprises, especially in the groups with high levels of competition, there is a strong positive correlation between the digital transformation of enterprises and the choice of diversification strategies, which indicates that market power is an important factor for enterprises to adopt diversification strategies in the context of digital transformation (Woodward et al., 2013), In other words, digital technology is not only a tool to maintain the monopoly position based on RBV theory, but also one of the important means and weapons to break through the monopoly market based on creative destruction theory.

For Hypothesis 4 and Hypothesis 5 are not true, generally speaking, from the perspective of corporate governance, the problems caused by agency problems and the holding of large shareholders and information asymmetry usually affect the strategic choice of enterprises (hautz et al., 2013; Denis et al., 1997). However, this study found that there was no significant difference in the group of large shareholders' shareholding ratio and information asymmetry. It shows that the issues of key shareholders' shareholding and information asymmetry are not the main considerations of diversification strategy selection in the process of enterprise digital transformation. This may be because although corporate governance and agency issues may affect the formulation of enterprise diversification strategy, in essence, the fundamental purpose of the formulation of diversification strategy is still around the main factors such as products, services, markets and projects(Kenny ,2009).

For the establishment of hypothesis 6, an important advantage of diversification strategy selection is that it can reduce the overall risk level of the enterprise (Gomez‐Mejia et al., 2010). This study found that in the group with high performance risk, the digital transformation of the enterprise can promote the diversification transformation of the enterprise, It shows that the risk of digital technology itself and the improvement of its risk bearing based on RBV theory are the main factors for enterprises to choose diversification strategies under the background of digital transformation.

With regard to the establishment of Hypothesis 7 and Hypothesis 8, we find that the influencing factors that cause enterprises to adopt diversification strategies in the context of digital transformation depend more on the characteristics and advantages of the diversification strategy itself, that is, to further obtain market power and reduce enterprise risks, while the traditional transaction cost and organizational cost factors that affect enterprise boundaries and corporate governance and agency issues do not play a major role.

## 14| CONCLUSIONS

This paper discusses the impact of digital transformation on enterprise diversification from both theoretical and empirical aspects for the first time. It is an important research paper in the field of digital research and enterprise diversification.Based on the above results, it is assumed that H3, H6 and H7 are valid, and H1, H2, H4, H5 and H8 are not. The study found that digital transformation can significantly promote the level of enterprise diversification, and the main conclusions of the study passed the robustness test and endogenous test. Cloud computing, digital technology application and big data play an important role in promoting enterprise diversification, while the impact of artificial intelligence and blockchain technology is not significant. We believe that the reason is that the former three technologies are more mature than the latter two.Through mechanism analysis, we find that the promotion effect of enterprise digital transformation on enterprise diversification is mainly realized through market power channel and firm risk channel, That is, the pursuit of establishing market power, monopoly profits and challenge the monopolistic position of market occupiers based on digital transformation and the decentralization strategy to deal with the risks associated with digital transformation are important reasons for enterprises to adopt diversification strategy under the background of digital transformation. Although the organization costs channel, transaction cost channel, blockholder control channel, industry type and information asymmetry channel have some influence on the main effect of this paper, they are not the main channel because they have not passed the inter group regression coefficient difference test statistically.

**IMPLICATION FOR MANAGEMENT**

This paper also has practical significance. By discussing the relationship between enterprise digital transformation and enterprise diversification strategy, this paper further expands the understanding of the economic impact of enterprise digital transformation. For policy makers, enterprise digital transformation is like a double-edged sword, it should pay close attention to the monopoly risk and systemic risk brought by the digital transformation of enterprises, and make good use of digital technology to break the market monopoly, appropriate incentive and restraint mechanisms shall be adopted for relevant subjects in the process of digital technology development, so as to support enterprises' digital innovation in the right track at the policy level. For enterprise managers, it must be recognized that the use of digital technology will bring more monopoly advantages to enterprises, and digital technology is also a powerful weapon to challenge market occupiers, and this advantage must be based on the good use of digital technology for diversification strategy to optimize product and factor combination to create more new value growth points. For the market occupiers, how to avoid the challenges of potential digital transformation enterprises is also a very noteworthy issue. At the same time, enterprise managers must also realize that there are certain technical risks in the use and application of digital technology, and enterprises must adopt a reasonable diversified mode to effectively share the risks, and it should also realize that the diversification strategy based on the digital transformation of enterprises has improved the level of enterprise risk-taking capacity. In addition, this paper finds that there are great differences between different digital technologies due to the maturity of the technology itself. Managers should pay attention to avoiding high-risk and immature technologies when making technology selection.

# APPENDIX A. VARIABLE DEFINITIONS

This table contains the definitions of variables used in our analysis.

| Total entropy | $$\text{Total entropy} = \sum_{i=1}^{N} S_i \ln(1/S_i)$$ where Si is the share of a firm's total sales in industry i and N is the number of industries in which the firm operates. . |
|---|---|
| Herfindahl index | $$\text{Herfindahl index} = \sum_{i=1}^{N} (S_i)^2$$ where Si is the share of a firm's total sales in industry i and N is the number of industries in which the firm operates. We take a negative number for this value for comparative analysis. |
| $Digitization_{i,t}$ | The Natural logarithm of 1 plus the total word frequency of digital keywords of company i in year t |
| $DTA_{i,t}$ | The Natural logarithm of 1 plus Digital technology application keyword word frequency of company i in year t |
| $AI_{i,t}$ | The Natural logarithm of 1 plus Artificial intelligence technology keyword word frequency of company i in year t |
| $BT_{i,t}$ | The Natural logarithm of 1 plus Blockchain technology keyword word frequency of company i in year t |
| $CC_{i,t}$ | The Natural logarithm of 1 plus Cloud computing technology keyword word frequency of company i in year t |
| $DT_{i,t}$ | The Natural logarithm of 1 plus Big data technology keyword word frequency of company i in year t |
| $Growth_{i,t}$ | current year's operating revenue / previous year's operating revenue - 1 of company i in year t |
| $Mshare_{i,t}$ | Total shares held by the management divided by the outstanding share capital of company i in year t. |
| $CI_{j,t}$ | This variable is measured by the ratio of industry j net fixed assets to industry j employees in year t |
| $HHI_{i,t}$ | The square sum of the companyi's operating income in the proportion of all companies in the industry in year t |
| $Soe_{i,t}$ | The value of state-owned holding enterprise is 1, and that of other enterprises is 0 |
| $Size_{i,t}$ | Natural logarithm of total assets of company i in year t |
| $Lev_{i,t}$ | Asset liability ratio = Total liabilities/total assets of company i in year t |
| $Roa_{i,t}$ | Return on total assets = Net profit/total assets of company i in year t |
| $Tophold_{i,t}$ | Shareholding ratio of the largest shareholder of company i in year t |
| $Dual_{i,t}$ | The dummy variable of management power of i company in year t, equal to 1 when the chairman and CEO are concurrently serving, |
| | otherwise it is 0. |
| $Maghold_{i,t}$ | The dummy variable of the management shareholding ratio of i company in year t; when the management shareholding ratio is greater than the annual industry median, it is 1, otherwise it is 0. |
| $FirmAge_{i,t}$ | Natural logarithm (current year - year of establishment + 1) of the company i in year t |
| $HHI2_{i,t}$ | The square sum of the companyi's Total asset in the proportion of all companies in the industry in year t |
| $BlockHolding_{i,t}$ | The total holding of block holders |
| $Analyst_{i,t}$ | Number of analysts following the firm. It is calculated by taking the natural logarithm of 1 plus the number of analysts following the firm at the end of a fiscal year |
| $Report_{i,t}$ | Number of research reports covering the firm. It is calculated by taking the natural logarithm of 1 plus the number of research reports issued by securities companies covering the firm in one fiscal year. |
| $risk1_{i,t}$ | $\sqrt{\frac{1}{T-1}\sum_{t=1}^{T}(adj-Roa_{i,t}-\frac{1}{T}\sum_{t=1}^{T}adj-Roa_{i,t})^2} \mid T=3$ |
| $risk2_{i,t}$ | $Max(adj-Roa_{i,t})-Min(adj-Roa_{i,t})$ |

## APPENDIX B. DIGITIZATION DEFINITIONS

| | **Digitization bag of words** | |
|---|---|---|
| | **Panel A. Bag of words (English Version)** | |
| | *type* | *Bag of words* |
| Bottom technology application | Artificial intelligence technology | Artificial intelligence, business intelligence, image understanding, investment decision support systems, intelligent data analysis, intelligent robots, machine learning, deep learning, semantic search, biometric technology, face recognition, voice recognition, identity verification, autonomous driving, natural language processing |
| | Big data technology | Big data, data mining, text mining, data visualization, heterogeneous data, credit investigation, augmented reality, mixed reality, virtual reality |
| | Cloud computing technology | Cloud computing, stream computing, graph computing, memory computing, multi-party secure computing, brain-inspired computing, green computing, cognitive computing, fusion architecture, billion-level concurrency, EB-level storage, Internet of Things, cyber-physical systems |
| | Blockchain technology | Blockchain, digital currency, distributed computing, differential privacy technology, smart financial contract |
| Technology practical application | Digital technology application | Mobile Internetwork, Industrial Internet, Mobile Internet, Internet Medical, E-commerce, Mobile Payment, Third Party Payment, NFC Payment, Smart Energy, B2B, B2C, C2B, C2C, O2O, Network Connection, Smart Wear, Smart Agriculture, Smart Transportation, Smart healthcare, smart customer service, smart home, smart investment advisory, smart cultural tourism, smart |

environmental protection, smart grid, smart marketing, digital marketing, unmanned retail, Internet finance, digital finance, Fintech, financial technology, quantitative finance, open banking

**Panel B. Bag of words (Chinese Version)**

| | type | Bag of words |
|---|---|---|
| 底层技术运用 | 人工智能技术 | 人工智能、商业智能、图像理解、投资决策辅助系统、智能数据分析、智能机器人、机器学习、深度学习、语义搜索、生物识别技术、人脸识别、语音识别、身份验证、自动驾驶、自然语言处理 |
| | 大数据技术 | 大数据、数据挖掘、文本挖掘、数据可视化、异构数据、征信、增强现实、混合现实、虚拟现实 |
| | 云计算技术 | 云计算、流计算、图计算、内存计算、多方安全计算、类脑计算、绿色计算、认知计算、融合架构、亿级并发、EB级存储、物联网、信息物理系统 |
| | 区块链技术 | 区块链、数字货币、分布式计算、差分隐私技术、智能金融合约 |
| 技术实践应用 | 数字技术运用 | 移动互联网、工业互联网、移动互联、互联网医疗、电子商务、移动支付、第三方支付、NFC支付、智能能源、B2B、B2C、C2B、C2C、O2O、网联、智能穿戴、智慧农业、智能交通、智能医疗、智能客服、智能家居、智能投顾、智能文旅、智能环保、智能电网、智能营销、数字营销、无人零售、互联网金融、数字金融、Fintech、金融科技、量化金融、开放银行 |

for strategic analysis. Strategic organization, 11(4), 403-411.